\documentclass[]{aa}
\usepackage{graphicx}
\usepackage{textcomp}
\usepackage{gensymb}
\usepackage[varg]{txfonts}
\usepackage{color}
\usepackage{minipage}
\usepackage{appendix}
\usepackage{multirow}
\usepackage{amsmath}
\usepackage{natbib}
\usepackage[colorlinks,linkcolor=blue,citecolor=blue,linktocpage=true,breaklinks,plainpages=false,urlcolor=blue]{hyperref}

\bibpunct{(}{)}{;}{a}{}{,}
\begin{document}
\title{Influence of NLTE effects in Fe~{\sc i} lines on  inverted atmosphere\\
II. 6301\,\AA{} and 6302\,\AA{} lines formed in 3D NLTE}
\author{H. N. Smitha \inst{1} \and R. Holzreuter \inst{1,2} \and M. van Noort \inst{1} \and S. K. Solanki \inst{1,3}}
\institute{$^{1}$Max-Planck-Institut f\"ur Sonnensystemforschung, 
Justus-von-Liebig-Weg 3, 37077 G\"ottingen, Germany\\
$^{2}$
Institute of Particle Physics and Astrophysics, ETH H\"onggerberg, 8093 Z\"{u}rich, Switzerland\\
$^{3}$School of Space Research, Kyung Hee University, Yongin, Gyeonggi, 446-701, Republic of Korea\\
\email{smitha@mps.mpg.de, solanki@mps.mpg.de}}
\titlerunning{3DNLTE effects in Fe~{\sc i} lines}
\authorrunning{Smitha et al.}

\abstract
{This paper forms the second part of our study on how the neglect of Non-Local Thermodynamic Equilibrium (Non-LTE, NLTE) conditions in the formation of Fe {\sc i} 6301.5\,\AA{} and the 6302.5\,\AA{} lines influences the atmosphere obtained by inverting their Stokes profiles in LTE. The main cause of NLTE effects in these lines is the line opacity deficit due to the excess ionization of the Fe~{\sc i} atoms by the ultraviolet (UV) photons in the Sun.}
{In the first paper, the above photospheric lines were assumed to have formed in 1DNLTE and the effects of horizontal radiation transfer (RT) were neglected. In the present paper, the iron lines are computed by solving the RT in 3D. We investigate the influence of horizontal RT on the inverted atmosphere and how it can enhance or reduce the errors due to the neglect of 1DNLTE effects.}
{The Stokes profiles of the iron lines are computed in LTE, 1DNLTE and 3DNLTE. They all are inverted using an LTE inversion code. The  atmosphere from the inversion of LTE profiles is taken as the reference model. The atmospheres from the inversion of 1DNLTE profiles (testmodel-1D) and 3DNLTE profiles (testmodel-3D) are compared with it. Differences between reference and testmodels are analysed and correspondingly attributed to NLTE and 3D effects.}
{The effects of horizontal RT are evident in regions surrounded by strong horizontal gradients in temperature. That is along the granule boundaries, regions surrounding magnetic elements and its boundaries with intergranular lanes. 
In some regions, the 3D effects enhance the 1DNLTE effects while in some, they weaken. In the small region analysed in this paper, the errors due to neglecting the 3D effects are less than $5\%$ in temperature. In a majority of the pixels, the errors are less than $20\%$ in both velocity and magnetic field strength. These errors are also found to survive when the Stokes profiles are spatially and spectrally degraded to the resolution of SST or DKIST.}
{The neglect of horizontal RT is found to introduce errors not only in the derived temperature but also in other atmospheric parameters. How large the errors are depends on how strong the local horizontal gradients are in temperature. Compared to the 1DNLTE effect, the 3D effects are more localised to specific regions in the atmosphere and overall less dominant. }
 
\keywords{Radiative transfer, Line: formation, Line: profiles, Sun: magnetic fields, Sun: photosphere, Polarization, Sun: atmosphere}
\maketitle

\section{Introduction}
\label{sec:intro}
The nature of radiative transfer problem for any stellar atmosphere diagnostic is inherently multi-dimensional and coupled with the effects of Non-local thermodynamic equilibrium (Non-LTE). For the solar atmosphere, one of the early studies where the two-dimensional (2D) radiative transfer problem was solved is in the construction of a spicule model by \cite{1969SoPh...10...88A} based on the Ca~{\sc ii} K line. This was followed by papers of \cite{1970ApJ...161..255C, 1971ApJ...169..157C, 1971JQSRT..11..559A, 1971SoPh...21...82C} among others, who solved the 2D transfer equation to study the solar chromosphere, derive solar atmospheric models and so on. \cite{1977A&A....54..577S, 1977A&A....58..273S, 1978A&A....67...33S} studied the effects of horizontal irradiation on Stokes profiles in magnetic flux tubes by solving the 2D transfer equation in cylindrical geometry. For a detailed review on this, see \cite{2008PhST..133a4012C}.

A few years later \cite{1985ASIC..152..215N} for the first time studied the spectral line formation in three-dimensions (3D) with NLTE effects in a realistic 3D model atmosphere. Since then, the 3DNLTE computations on 3D model atmospheres have been used to compute spectral lines of, for example, lithium \citep{2003A&A...399L..31A}, oxygen \citep{1995A&A...302..578K, 2004A&A...417..751A, 2005A&A...435..339A, 2009A&A...508.1403P, 2013MSAIS..24..111P, 2015A&A...583A..57S}, iron \citep{2017MNRAS.468.4311L}, manganese \citep{2019A&A...631A..80B}, barium \citep{2020A&A...634A..55G}, for the determination of their abundances in solar and other stellar atmospheres. Recently, forward modeling of Chromospheric lines such as the Ca~{\sc ii} 8542\,\AA{} \citep{2009ApJ...694L.128L}, Na~{\sc i} D1 line \citep{2010ApJ...709.1362L}, H$\alpha$ \citep{2012ApJ...749..136L, 2015ApJ...802..136L}, Mg~{\sc ii} h \& k lines \citep{2013ApJ...772...90L}, and the Ca~{\sc ii} H \& K lines \citep{2013ApJ...767..108A, 2018A&A...611A..62B} were carried out in 3DNLTE to understand the line formation and to study different chromospheric features observed with these lines.  \cite{2017A&A...597A..46S} synthesized the Mg~{\sc ii} h \& k, Ly$\alpha$ and Ly$\beta$ lines taking into account the complexities of partial frequency redistribution in 3DNLTE. \cite{2019A&A...631A..80B} modeled different chromospheric lines observed in active regions using 3DNLTE computations.  

\begin{figure*}
    \centering
    \includegraphics[width=0.28\textwidth]{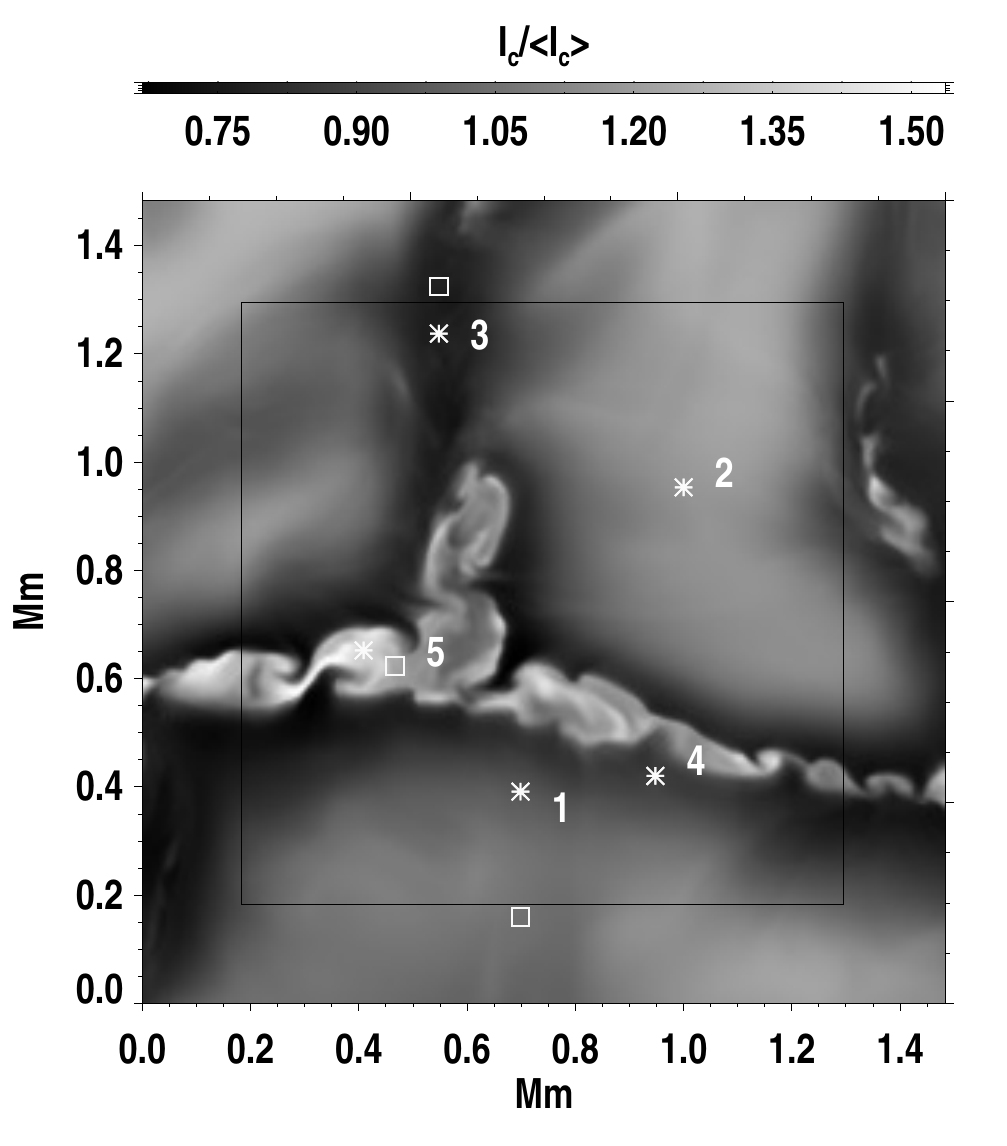}
    \includegraphics[width=0.70\textwidth]{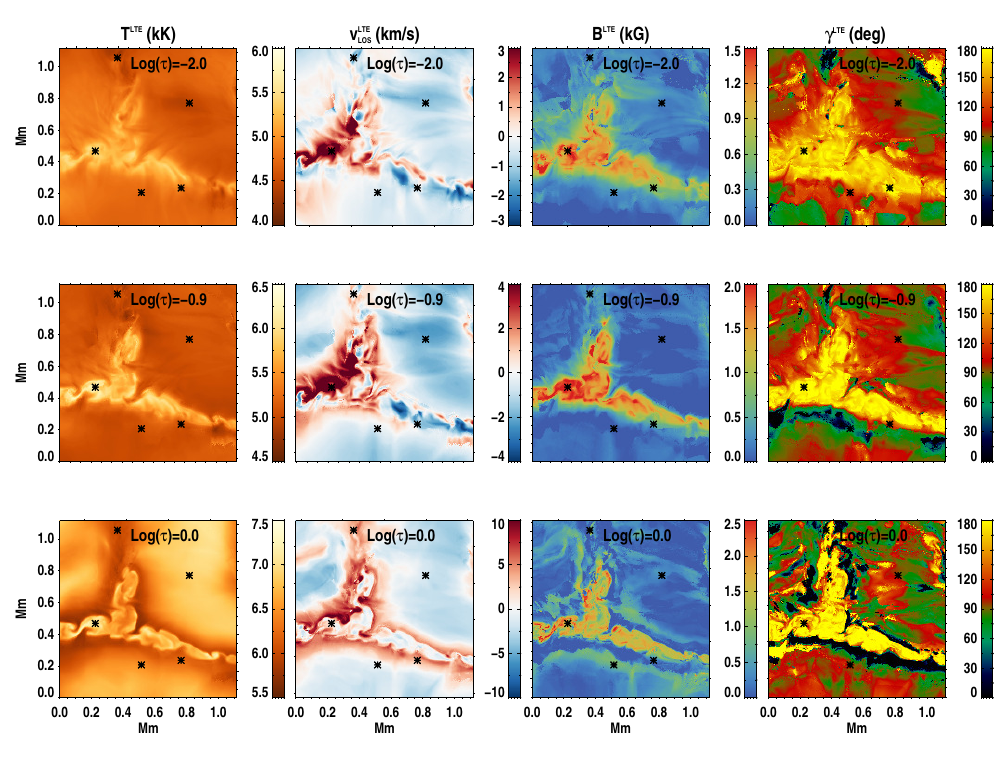}
    \caption{\textit{Left}: Continuum intensity image normalised with spatially averaged continuum intensity. It is a small patch of a bigger cube used in paper I. The inner black box indicates the actual area used for analysis in the rest of this paper. The area outside of this box has been neglected because of the minor uncertainties in the intensity profiles which can result from the non-periodic boundary conditions. The five representative spatial points are marked. For pos.\,1, pos.\,3 and pos.\,5, their old locations used in paper I are indicated by white squares.  \textit{Right}: Maps of temperature ($T$), LOS velocity ($v_{\rm LOS}$) and magnetic field strength ($B$), at $log(\tau)=-2.0, -0.9$ and $0.0$, obtained from the inversion of LTE profiles. This will be used as the "reference model". In all the 2D maps in this paper, the origin is taken as (0,0).}
    \label{fig:maptau}
\end{figure*}

\begin{figure*}[]
\centering
\includegraphics[width=\textwidth]{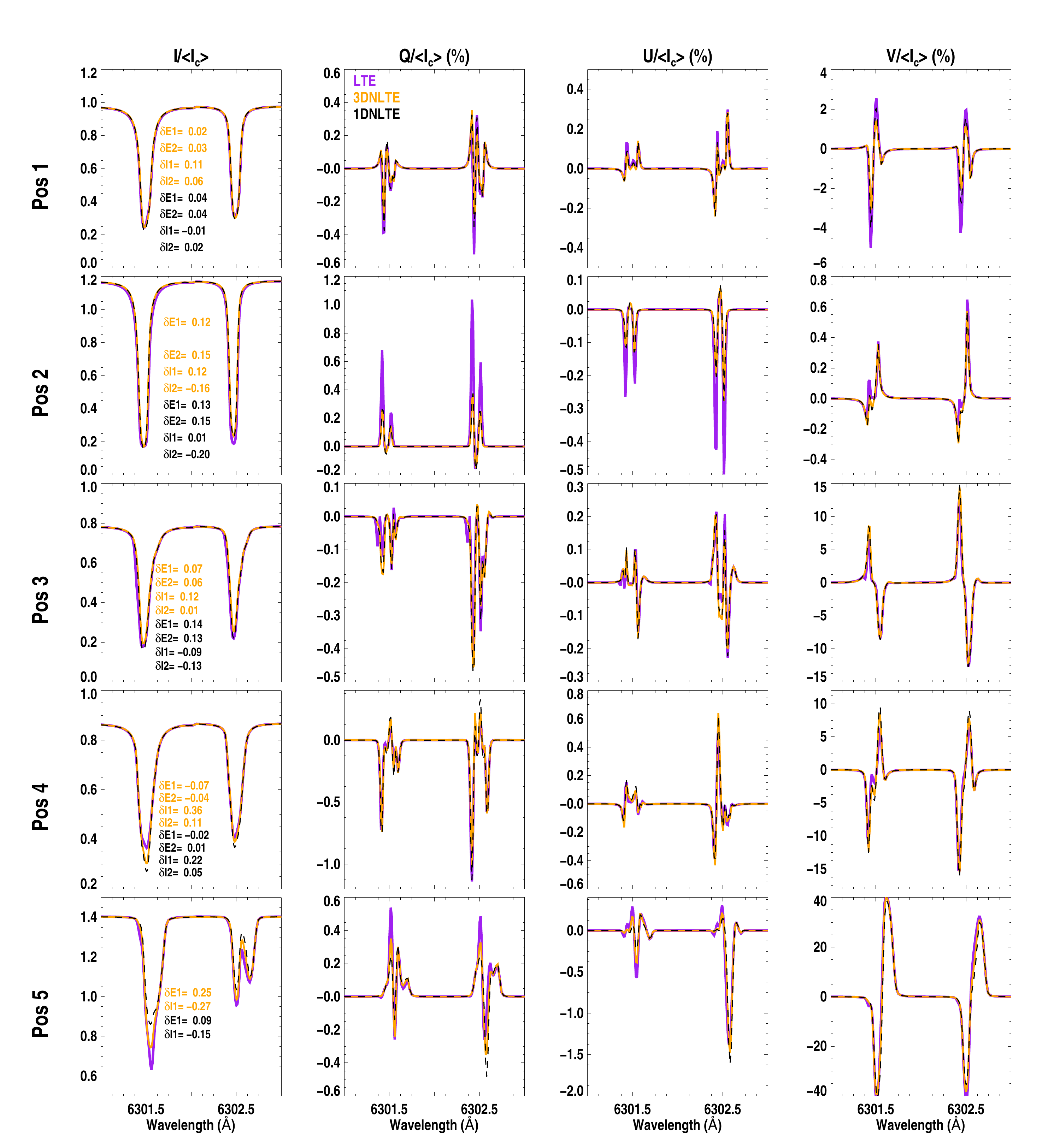}
\caption{The Stokes profiles at the representative five spatial points indicated in Figure~\ref{fig:maptau}. The profiles computed in LTE, 1DNLTE and 3DNLTE are indicated in purple, black and orange respectively. In the first column, the differences in equivalent widths and the residual intensities with respect to the LTE profiles are indicated for the two lines. They are computed using Equation~\ref{eqn:int_ew}. For better visibility of the differences between the profiles, we again show the 1DNLTE and 3DNLTE profiles separately in figure~\ref{fig:profiles2}. }
\label{fig:profiles}
\end{figure*}

In the present paper, we focus on the 3DNLTE effects in the  Fe~{\sc i} 6301.5\,\AA{} and 6302.5\,\AA{} lines formed in a network region. Although their line core formation height is in the upper photosphere, they are affected by the NLTE conditions in the atmosphere. This is because of the over-ionization of the Fe~{\sc i} atoms by the ultra-violet (UV) photons. This underpopulation of Fe~{\sc i} atoms results in line opacity deficit and lowers the departure coefficient to a value below unity. Due to their formation height, the lines are also affected by the source function deficit. Depending on which of two effects dominates, the NLTE lines are either weaker or stronger than the LTE lines. These effects have now been known for more than four decades and are discussed in the papers by \citet{1972ApJ...176..809A,1982A&A...115..104R, 1988ASSL..138..185R, 1988A&A...189..243S, 2001ApJ...550..970S}, to name a few. In all these papers, the radiative transfer equation was solved in one-dimension (1D) and the effects of multi-dimensions were neglected.  In a series of papers, \cite{2012A&A...547A..46H, 2013A&A...558A..20H, 2015A&A...582A.101H} for the first time computed the photospheric iron lines at 5247\,\AA{}, 5250\,\AA{}, 6301.5\,\AA{} and 6302.5\,\AA{} in 3DNLTE using different model atmospheres, namely, a simple flux tube model, 3D hydrodynamic (HD) simulation and 3D Magnetohydrodynamic (MHD) simulation. They presented a detailed comparison of the Stokes profiles computed in LTE, 1DNLTE and 3DNLTE and highlighted the importance of accounting for the 3DNLTE effects in these lines. However, no comparison with observations was made. \cite{2017MNRAS.468.4311L} modelled the observed centre-to-limb variation in intensity for many iron lines, different from the above four, by solving the 3DNLTE radiative transfer and measured the solar iron abundance.

Since the  Fe~{\sc i} 6301.5\,\AA{} and 6302.5\,\AA{} lines are extensively used for solar photospheric diagnostics, it is important to account for the multi-dimensional NLTE effects in their analysis. The routinely employed Stokes profiles inversions assume that the lines are formed in LTE, which is always 1D in diagnostic radiative transfer.  Neglecting the 3DNLTE effects introduces errors in the derived atmospheric model. Understanding by how much the 3DNLTE effects influence the inverted atmosphere is the main aim of the present study. In the first paper of this series, we assumed the iron lines to be formed in 1DNLTE conditions and estimated the errors in the atmosphere derived by inverting their Stokes profiles using an LTE inversion code \citep[][hereafter paper I]{2020A&A...633A...157S}. We found that neglecting the 1DNLTE effects introduces errors in the measurement of all atmospheric parameters. In temperature the errors can be as large as 13\% and in the line-of-sight (LOS) velocity, magnetic field strength, the errors can go upto 50\% or more. Even the measurement of magnetic field inclination is prone to errors when the 1DNLTE effects are neglected. These effects were evident in Stokes profiles at granules, intergranular lanes, magnetic elements, their boundaries, basically in every region with strong vertical gradients in temperature, LOS velocity or magnetic field. 

In the present second paper of this series, we account for the horizontal radiative transfer (RT) effects in the iron lines and investigate how they influence the inverted atmosphere. Similar to paper I, we compute the Stokes profiles in LTE and 3DNLTE conditions, and invert them using an LTE inversion code. The  LTE inversion of LTE profiles is self-consistent and we treat the derived atmospheric model as a reference. The inverted atmosphere from the LTE inversion of 3DNLTE profiles is not self-consistent, so that any departures with respect to the reference model can be attributed to the 3DNLTE effects. More details on this approach will be discussed in the sections below. As discussed in \citep{2012A&A...547A..46H, 2013A&A...558A..20H, 2015A&A...582A.101H}, the horizontal RT effects can either weaken or strengthen the 1DNLTE effects. In the present paper, we investigate by how much the 3D effects strengthen/weaken the 1DNLTE effects and thus increase/decrease the errors in the inverted atmosphere.

\section{Radiative transfer computations} 
\subsection{Stokes profiles synthesis}
\label{sec:atmos}
The Stokes profiles of the Fe~{\sc i} lines at 6301.5\,\AA{} and 6302.5\,\AA{} were synthesized in LTE, 1DNLTE and 3DNLTE using a modified version of the RH code \citep{2001ApJ...557..389U}. The modifications are explained in \cite{2012A&A...547A..46H}. The details of the computational setup for the 3DNLTE run are described in \citet{2013A&A...558A..20H, 2015A&A...582A.101H}. The snapshot of the MHD simulation used for the synthesis is the same as the one used in paper I. It was generated using the MuRaM code \citep{2005A&A...429..335V} and has a grid spacing of $5.82$\,km in the $xy-$direction and $7.85$\,km in the vertical direction. Since solving the radiative transfer equation in 3D is computationally expensive and time consuming, the line profiles were synthesized over a smaller region shown in Figure~\ref{fig:maptau}. In the rest of the paper, we refer to this smaller region as simply, the "sub-region". It comprises a kilo-Gauss strength magnetic structure embedded in an intergranular lane and surrounded by sections of three granules. The continuum image in Figure~\ref{fig:maptau} shows strong horizontal gradients in the intensity, an indicator of the susceptibility of this region to 3D RT effects \citep{2013A&A...558A..20H}.

We use an Fe~{\sc i} atomic model with 23 levels coupled by 33 line transitions and 22 bound-free transitions. The missing UV opacity was treated using a fudge factor \citep{1972ApJ...176..809A}. The other details of the model atom and profile synthesis are described in \citet{2013A&A...558A..20H,2015A&A...582A.101H} and in paper I. 

\subsection{Inversion of Stokes profiles}
\label{sce:inversion}
The Stokes profiles computed in LTE, 1DNLTE and 3DNLTE were inverted in LTE using the SPINOR code \citep{1987PhDT.......251S,2000A&A...358.1109F}. The code returns a simple atmospheric model with five parameters, temperature ($T$), line-of-sight velocity ($v_{\rm LOS}$), and magnetic field vector ($B,\gamma,\chi$), at optical depths, $log(\tau)=0.0, -0.9$ and $-2.0$ corresponding to the chosen three inversion nodes. They were chosen based on the optimum placement discussed in \citet{2016A&A...593A..93D}. The $log(\tau)$ scale is constructed at a reference wavelength of 5000\,\AA. Further details on the inversion can be found in paper I. 

For the normalization of the Stokes profiles, we use the spatially averaged continuum intensity ($\langle I_c \rangle$). Since the 3D NLTE profiles are computed only over the sub-region, the $\langle I_c \rangle$ is different from the $\langle I_c\rangle$, that was used for the normalization of the LTE and 1DNLTE profiles, which were computed over the whole atmospheric cube, as described in paper I. For a consistent comparison with the 3DNLTE profiles, we have therefore repeated the inversion of LTE and 1DNLTE profiles over the sub-region using the $\langle I_c \rangle$ computed over that region. 

\begin{figure*}
\centering
\includegraphics[width=\textwidth]{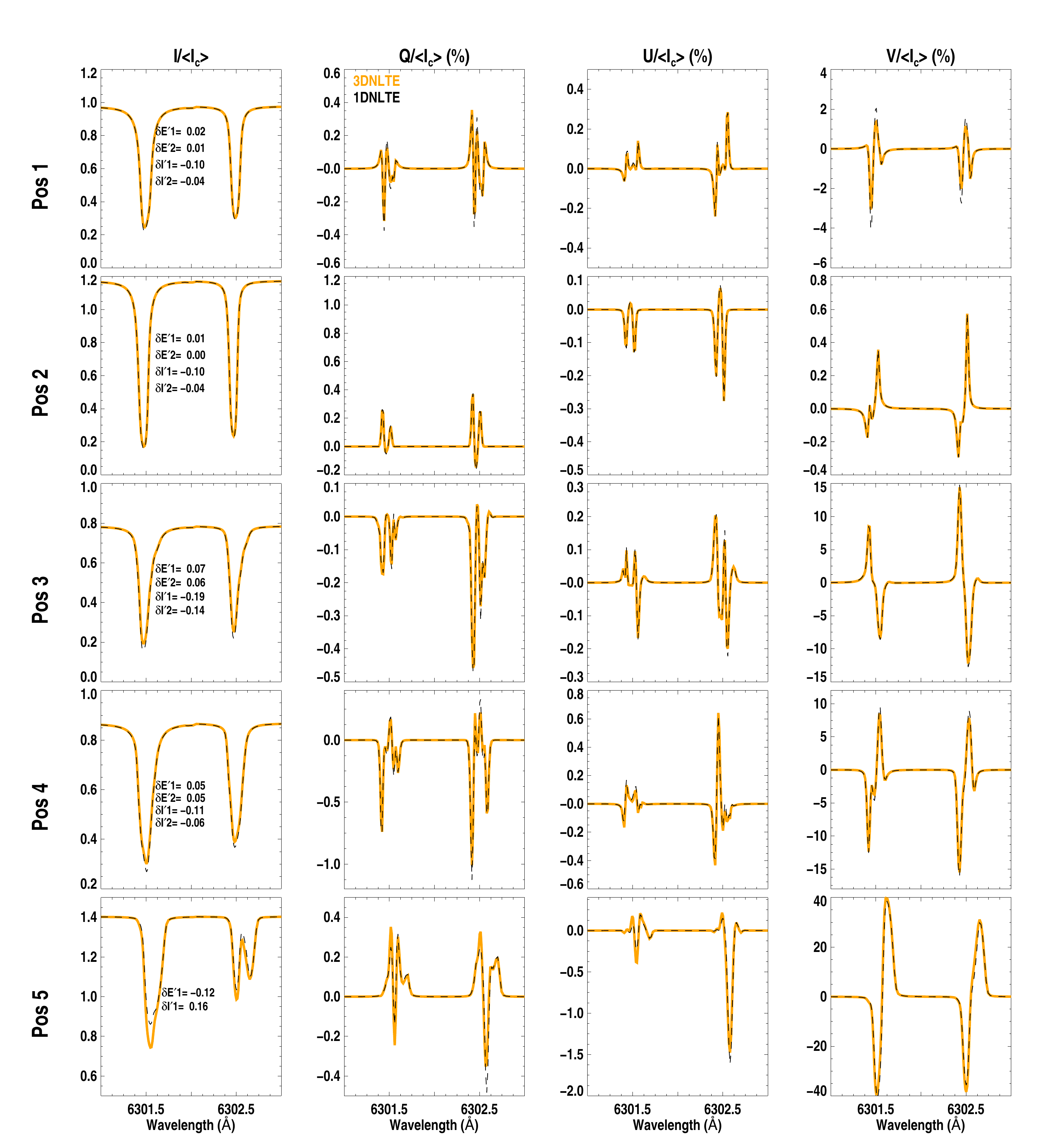}
\caption{Similar to Figure~\ref{fig:profiles} but showing a comparison between the profiles computed in 1DNLTE (black) and 3DNLTE (orange). The {$\delta E^{\prime}$ and $\delta{I^{\prime}}$ measure the difference in equivalent widths and residual intensities between the 1DNLTE and 3DNLTE profiles. They are computed using Equation.~\ref{eqn:int_ew_2}}.}
\label{fig:profiles2}
\end{figure*}

\section{Comparison of different atmospheres}
\label{sec:comp}
To compare the different inverted atmospheres and to investigate the influence of NLTE and 3D RT effects on the atmospheric parameters, we use a strategy similar to that used in paper I. We refer to the atmosphere obtained by inverting LTE profiles as the "reference model". The atmospheres inferred from the inversion of 1DNLTE profiles and 3DNLTE profiles are referred to as "testmodel-1D" and "testmodel-3D", respectively. The reference model is used as the fiducial model to assess the amount of change in the two testmodels. Maps of $T, v_{\rm LOS}, B$ and $\gamma$ at the three inversion nodes $log(\tau)=0.0, -0.9$ and $-2.0$ in the reference model are shown in Figure~\ref{fig:maptau}. 

The sub-region chosen for the analysis of 3D effects contains structures with a wide range of temperature, velocity and magnetic field values, which in addition vary with height in the atmosphere. While the stronger velocity fields and magnetic fields can be determined with higher accuracy, the weak fields are quite difficult to constrain, which results in uncertainties in the inverted atmosphere. The 3-node atmospheric model used here for the inversions, though typical of the more complex models used to invert observations,  is an over-simplification of the complex MHD cube used for the spectral synthesis. Since there is no uniquely correct way to compare this 3-node atmosphere with the actual MHD cube, we believe there is no easy way to quantify the accuracy of either the reference or the test models, against the atmosphere used for synthesis. As discussed also in paper I, this is the reason why we compare the testmodels with a reference model atmosphere which also is a 3 node atmosphere obtained from inversions and hence prone to the same uncertainties, except for those introduced by neglecting NLTE or 3D effects. Such a choice is expected to alleviate the contributions from the inversion uncertainties in our analysis. In addition, the results presented below come from a statistical study to isolate the features/locations in the atmosphere where one can expect 1DNLTE and or 3D transfer effects, and to estimate how much they can influence the inverted atmosphere. Hence, we assume that any departures in testmodel-1D are purely due to 1DNLTE effects, while the deviations in the testmodel-3D are the result of a combination of 1DNLTE and 3D RT effects. In order to isolate the 3D RT effects from the 1DNLTE effects, we compare testmodel-3D with testmodel-1D. The differences between them are attributed to horizontal RT effects. 

Similar to paper I, we compute the simple and relative differences between the atmospheric models to compare each parameter. We define
\begin{eqnarray}
    \Delta x^{\rm \sc{nD}} = x^{\rm \sc{LTE}}-x^{\rm \sc{nDNLTE}};
    \delta x^{\rm nD} = \frac{x^{\rm LTE}-x^{\rm nDNLTE}}{x^{\rm LTE}},
    \label{eqn:rel_diff1}
\end{eqnarray}
where $\Delta x$ and $\delta x$ represent the simple and relative differences respectively. $x$ is any atmospheric quantity, such as $T, v_{\rm LOS}, B, \gamma$ and $nD$ stands for either 1D or 3D. To compare the two testmodels, we use
\begin{eqnarray}
\centering
    \Delta x^{\rm \sc{3D}}_{\rm \sc{1D}} = x^{\rm \sc{1DNLTE}}-x^{\rm \sc{3DNLTE}};
    \delta x^{\rm \sc{3D}}_{\rm \sc{1D}} = \frac{x^{\rm \sc{1DNLTE}}-x^{\rm \sc{3DNLTE}}}{x^{\rm \sc{1DNLTE}}},
    \label{eqn:rel_diff2}
    \end{eqnarray}
Since the atmospheric maps of the two test models look very similar to the reference models, we only show the difference maps in the following sections. However, the maps of all the atmospheric quantities from testmodel-1D and testmodel-3D are included in the Appendix~\ref{sec:appendix-a}.

After a statistical comparison between the different atmospheric models, we go into the details of individual features sampled by 5 representative spatial positions. Pos.\, 2, and pos.\, 4 are identical to those used in paper I. However, Pos.\,1 and pos.\,3 are moved in the $y$-direction to bring them into the black box indicated in the continuum image in Figure~\ref{fig:maptau}. This black box indicates the actual region that will be used for analysis in the rest of the paper. In all the figures, only the region within this box is displayed. The region outside of this box has been neglected due to the unphysical behaviour of the intensity profiles, which is a result of the non-periodic boundary conditions. This was identified by performing two 3D calculations on different but overlapping areas of the simulation box. Relative differences between intensities and Equivalent Width (EW) of the profiles in this overlapping area were compared. It was found that the influence of the non-periodic boundary is restricted to the pixels near the boundary of a computed domain. By removing the 32 rows/lines of pixels nearest to each boundary, it could be ensured that the influence of the boundary on the profiles was smaller than $1\%$ everywhere. More details on this are described in Holzreuter \& Solanki (in prep.). The new pos.\,1 lies closer to the edge of the granule, and pos.\,3 within the same intergranular lane. Pos.\,5 has also been shifted to a region which shows enhanced 3D effects, but it is still within the same magnetic structure. All the original and shifted spatial positions are marked in Figure~\ref{fig:maptau}.

In Figure~\ref{fig:profiles}, we compare the Stokes profiles from the LTE, 1DNLTE, and 3DNLTE runs. For clarity, we compare just the 1DNLTE and 3DNLTE profiles in a separate Figure~\ref{fig:profiles2}. The differences in their equivalent widths and residual intensities are computed using
 \citep[see also][]{2015A&A...582A.101H, 2020A&A...633A...157S}
\begin{eqnarray}
    \delta I = \frac{I^{\rm LTE}-I^{\rm nDNLTE}}{I^{\rm nDNLTE}};
     \delta E = \frac{EW^{\rm LTE}-EW^{\rm nDNLTE}}{EW^{\rm nDNLTE}},
    \label{eqn:int_ew}
\end{eqnarray}
where $I^{\rm LTE}, I^{\rm nDNLTE}$ is the minimum intensity of an LTE/nDNLTE profile at a specific spatial location and ${\rm nD}$ stands for either 1D or 3D. The equivalent widths are represented by $EW^{\rm LTE}, EW^{\rm nDNLTE}$. {To compare the 1DNLTE and 3DNLTE profiles in Figures~\ref{fig:profiles2} and \ref{fig:prof_specmear}, we use,}
\begin{eqnarray}
    \delta I^{\prime} = \frac{I^{\rm 1DNLTE}-I^{\rm 3DNLTE}}{I^{\rm 3DNLTE}};
     \delta E^{\prime} = \frac{EW^{\rm 1DNLTE}-EW^{\rm 3DNLTE}}{EW^{\rm 3DNLTE}}.
    \label{eqn:int_ew_2}
\end{eqnarray}

{The LTE fit to the Stokes $I$ and $V$ profiles computed in 1DNLTE and 3DNLTE at the five representative positions are shown in Figure~\ref{fig:fit_profiles}. }

\begin{figure*}
\centering
\includegraphics[width=\textwidth]{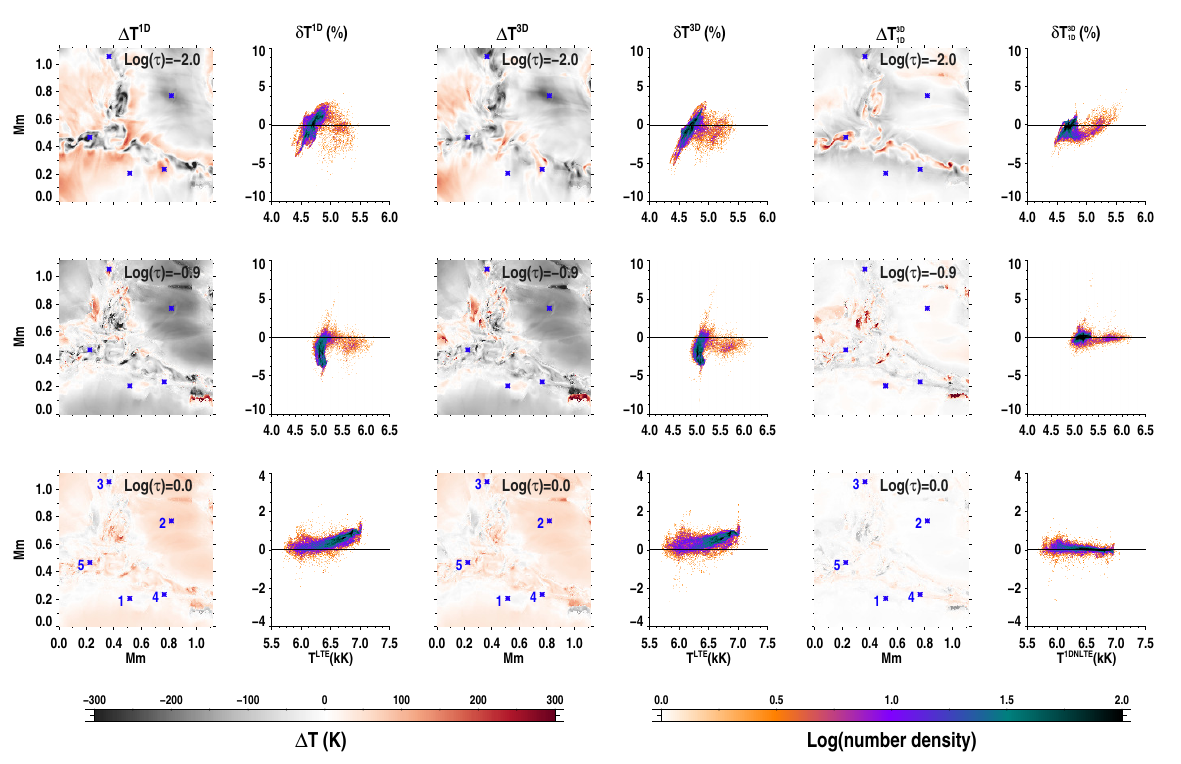}
\caption{Difference in the temperature maps from the reference model and testmodel-1D (\textit{first column}), reference model and testmodel-3D  (\textit{third column}). The \textit{fifth column} is the difference between the temperatures in testmodel-1D and testmodel-3D. The scatter density plots next to each difference image is the corresponding relative differences, defined in Equations~\ref{eqn:rel_diff1} and \ref{eqn:rel_diff2}. 
See Section~\ref{sec:temp} for more details.}
\label{fig:temp}
\end{figure*}

\begin{table}[]
    \centering
    \begin{tabular}{r r r r}
     & $\Delta T^{\rm 1D}$ (K) &$\Delta T ^{\rm 3D}$ (K) & $\Delta T^{\rm 3D}_{\rm 1D}$ (K) \\
\vspace{2pt}
    ${log(\tau)}$ & \textit{Mean, S.D} & \textit{Mean, S.D}. & \textit{Mean, S.D.}\\
    ${-2.0}$ &$-10.14,\ 71.25$ & $-31.40,\ 64.41$ &$ -21.27,\ 43.24$\\
    ${-0.9}$ &$-75.39,\ 71.11$ & $-75.89,\ 70.80$ & $-0.50,\ 36.04$\\
    ${0.0}$ & $27.95,\ 27.43$ & $27.91,\ 27.62$ & $-0.03,\ 15.79$\\
%
        & & & \\
             \hline
       & & & \\
\vspace{2pt}
  & $\Delta v^{\rm 1D}$ (km/s) &$\Delta v ^{\rm 3D}$ (km/s) & $\Delta v^{\rm 3D}_{\rm 1D}$ (km/s) \\
         \vspace{2pt}
     ${log(\tau)}$ & \textit{Mean, S.D} & \textit{Mean, S.D}. & \textit{Mean, S.D.}\\
     ${-2.0}$ &$-0.16,\ 0.26$ & $-0.17,\ 0.25$ & $-0.01,\ 0.13$\\
     ${-0.9}$ &$-0.00,\ 0.31$ &  $0.00,\ 0.33$ &  $0.01,\ 0.26$\\
     ${0.0}$ &$0.01,\ 0.49$& $-0.01,\ 0.48$ & $-0.02,\ 0.36$\\
      %
    & & & \\
    \hline
       & & & \\
         \vspace{2pt}
     & $\Delta B^{\rm 1D}$ (G) &$\Delta B ^{\rm 3D}$ (G) & $\Delta B^{\rm 3D}_{\rm 1D}$ (G) \\
       \vspace{2pt}
 ${log(\tau)}$ & \textit{Mean, S.D} & \textit{Mean, S.D}. & \textit{Mean, S.D.}\\ 
     ${-2.0}$& $22.28,\ 142.01$ & $42.73,\ 133.23$ & $20.46,\ 85.64$\\
     ${-0.9}$ &$1.05,\ 105.00$ & $ 5.45,\ 115.62$ & $4.39,\ 80.42$\\
    ${0.0}$& $-22.53,\ 216.33$  & $-5.00,\  215.94$ & $17.53,\ 176.19$\\
 %
    & & & \\  
            \hline
    & & & \\
        \vspace{2pt}
     & $\Delta \gamma^{\rm 1D}$(deg) &$\Delta \gamma^{\rm 3D}$ (deg) & $\Delta \gamma^{\rm 3D}_{\rm 1D}$ (deg) \\
     \vspace{2pt}
${log(\tau)}$& \textit{Mean, S.D} & \textit{Mean, S.D}. & \textit{Mean, S.D.}\\
    ${-2.0}$ &$-1.78,\ 23.29$ & $-1.73,\ 24.69$ & $0.04,\ 16.77$\\
     ${-0.9}$ &$ 0.74,\ 16.11$ & $-0.21,\ 19.56$ & $-0.95,\ 15.96$\\
     ${0.0}$ &$1.41,\ 26.40$ & $0.79,\ 28.69$  & $-0.62,\  20.33$\\
    & & & \\

    \end{tabular}
    \caption{The above table summarizes differences between reference model, testmodel-1D and testmodel-3D for the four atmospheric parameters. See Section~\ref{sec:temp} for explanations. 
    }
    \label{tab:table1}
\end{table}

\subsection{Temperature}
\label{sec:temp}
A comparison between the temperatures in the reference model and the two testmodels at the three nodes is shown in Figure~\ref{fig:temp}. Although this case has already been discussed in detail in paper I, maps of the simple difference ($\Delta T^{\rm 1D}$) and relative difference ($\delta T^{\rm 1D}$) of the reference model and the testmodel-1D are shown here for convenience (\textit{first and second columns}). In the central and bottom nodes, the $\Delta T^{\rm 3D}$ and $\delta T^{\rm 3D}$ maps, comparing the testmodel-3D with the reference model (\textit{third and fourth columns}), are quite similar to $\Delta T^{\rm 1D}$ and $\delta T^{\rm 1D}$. Statistically, both testmodel-1D and testmodel-3D differ from the reference model by up to 5\% in the top and central nodes. This is much smaller than the 13\% quoted in paper I because in the present paper, we analyse only a small sub-region of the full cube used in paper I.  

In the bottom node, the differences are non-zero and go up to 2\% (50 k -- 100k), similar to what was observed also in paper I. The reason for this is that within the spectral window used for the synthesis of line profiles, the real continuum has not been reached. In addition, the information from $log(\tau)=0.0$ is partly transported through the NLTE opacity of the line, and thus the maps at $log(\tau)=0.0$ are different. Although the continuum is computed in LTE for all the three runs (LTE, 1DNLTE and 3DNLTE), one can expect to recover the same inverted temperature in the bottom node only if a significant fraction of the spectral range has true continuum in it. For the same reason, we also have non-zero $\Delta T^{\rm 3D}$ at $log(\tau)=0.0$.

In Table~\ref{tab:table1}, we summarize the differences between the reference model, testmodel-1D and testmodel-3D. For each of the four parameters, $T, v_{\rm LOS}, B$ and $\gamma$, we first compute the simple difference $\Delta$ defined in Equation~\ref{eqn:rel_diff1}. Then we plot a histogram for each case (not shown in this paper but similar histograms for the 1DNLTE case are shown in paper I). The mean and standard deviations (S.D.) of the histogram distribution for all the four atmospheric parameters at the three inversion nodes are given in this table. The mean and S.D. gives us an estimate of how much the two testmodels differ from the reference model and also from each other.

For the temperature, the mean and S.D. for the distribution $\Delta T^{\rm 3D}$ computed for the testmodel-3D is less than 100 K in all the three nodes and comparable to the mean and S.D in the $\Delta T^{\rm 1D}$ computed for the testmodel-1D. However in the top node, we see an increase in the mean value of errors in testmodel-3D. This means that over the sub-region considered here, on an average, the 3D effects go in the same direction as the 1D NLTE effects, making the difference to the reference model larger.

 To isolate the 3D effects from the 1DNLTE effects, we compare testmodel-1D and testmodel-3D using Equations~\ref{eqn:rel_diff2} and plot them in the last two columns of Figure~\ref{fig:temp}. Of the three nodes, {differences as large as 300\,K} are seen only in the top node. The mean of $\Delta T^{\rm 3D}_{\rm 1D}$ histogram is also the largest in this node, {21\,K and 43\,K, respectively} (Table~\ref{tab:table1}). Differences are seen along the strong magnetic element, the intergranular lane surrounding it, and also in the granule-intergranular lane boundaries. These are the regions with a strong horizontal temperature gradient. In the top node, the $\Delta T^{\rm 3D}_{\rm 1D}$ is positive along the magnetic structure (red color). That is, the temperature in the testmodel-1D is higher than in testmodel-3D. This is because of the horizontal irradiation from the cooler surroundings, which makes the lines computed in 3DNLTE stronger than in 1DNLTE, as also discussed in \cite{2015A&A...582A.101H}. For example, the intensity profiles at Pos.\,5 plotted in Figure~\ref{fig:profiles}, which lies within this magnetic structure, clearly shows this effect. At this pixel, the 3DNLTE profile is stronger than the 1DNLTE effects. But the $\delta E$ and $\delta I$ are smaller for the 3DNLTE lines than the 1DNLTE lines.  The 3D effects produce changes in the opposite direction to the 1DNLTE effects. Thus, at $log(\tau=-2.0)$, the temperature in the testmodel-3D is closer to the reference model than in the testmodel-1D, see Figure~\ref{fig:1d-3d}.

In Pos.\,3, which is in an intergranular lane, the 3D effects weaken the intensity profiles (see Figure~\ref{fig:profiles}). This is because of the irradiation from the hotter surroundings. At this pixel, the temperature in the testmodel-3D is higher than in both testmodel-1D and reference model at $log(\tau)=-2.0$, see Figure~\ref{fig:1d-3d}. This is an example of one such pixel where the 3D effects further enhance the NLTE effects and neglecting them both will introduce combined error as large as 250 K in the temperature at the top node. 

Another interesting feature, represented by Pos.\, 4, is the intergranular lane right next to a bright magnetic structure. It is subject to hot irradiation from the magnetic structure on one side and a granule from the other side. Accordingly, the 3DNLTE intensity profiles of the two lines are weaker than the 1DNLTE profiles. The differences in residual intensities between 3DNLTE and LTE runs have decreased. The testmodel-3D returns a $T(log(\tau)=-2.0)$ value closer to that found by the reference model (see Figure~\ref{fig:1d-3d}). 
  
The temperatures at pos.\,1 and pos.\,2  are much less affected by 3D effects compared to the other three positions discussed above. At Pos.\, 1, we see that the 3DNLTE profiles are slightly weaker than the 1DNLTE profiles, whereas at Pos.\, 2, they are nearly the same. This can be explained based on the nature of the surrounding environment, in a way similar to the other three cases discussed above.
    
    In the above five representative pixels, the 3DNLTE effects, in intensity, are seen to mostly affect the depth of the line profile, which results in differences in temperature between different models in the top node. In the central node, the $\Delta T^{\rm 3D}_{\rm 1D}$ are quite small and are observed to have significant values only in a few pixels close to the magnetic structure.  The $\delta T^{\rm 3D}_{\rm 1D}$ is mostly less than 1\%. The line wings are much less affected by the 3D effects. In the bottom node, the $\Delta T^{\rm 3D}_{\rm 1D}$ and $\delta T^{\rm 3D}_{\rm 1D}$ is nearly zero. The mean of the histogram distributions are also nearly zero in both the central and bottom nodes.  

\begin{figure*}
\centering
\includegraphics[width=\textwidth]{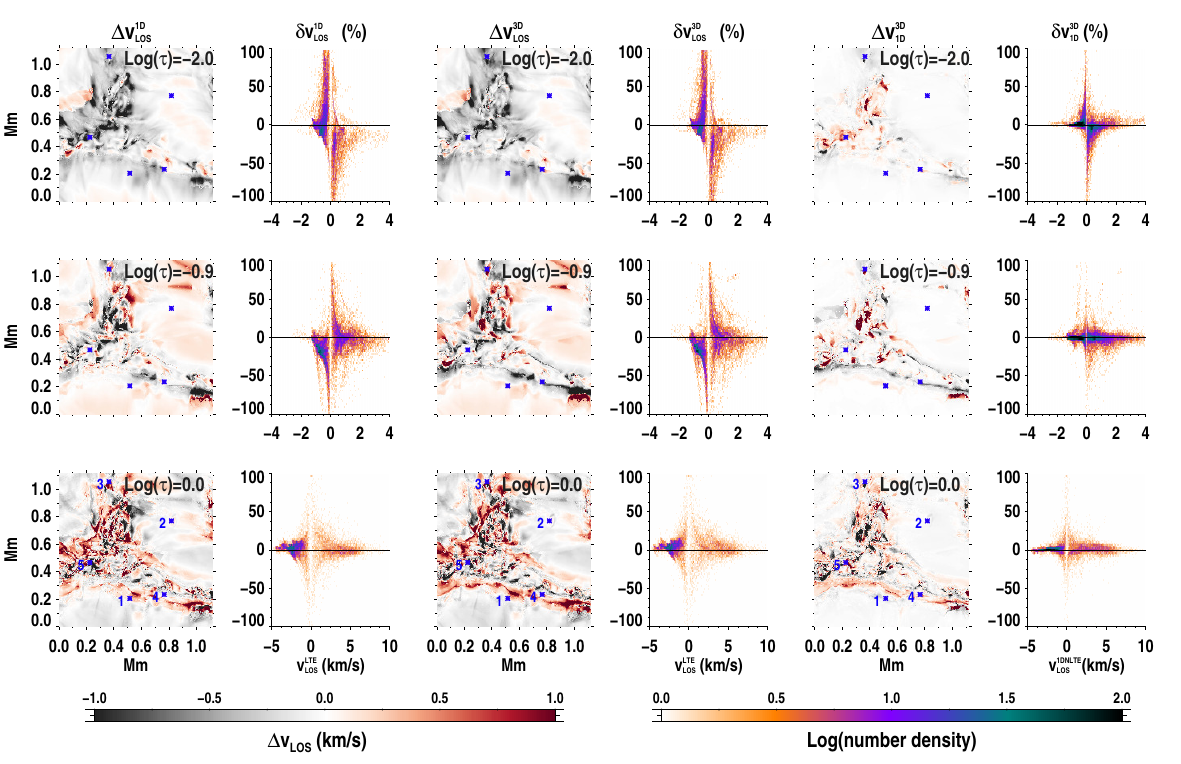}
\caption{Same as Figure~\ref{fig:temp} but for line-of-sight velocity. See Section~\ref{sec:vlos} for further details.}
\label{fig:vlos}
\end{figure*}

\begin{figure*}[htbp]
\centering
\includegraphics[width=\textwidth]{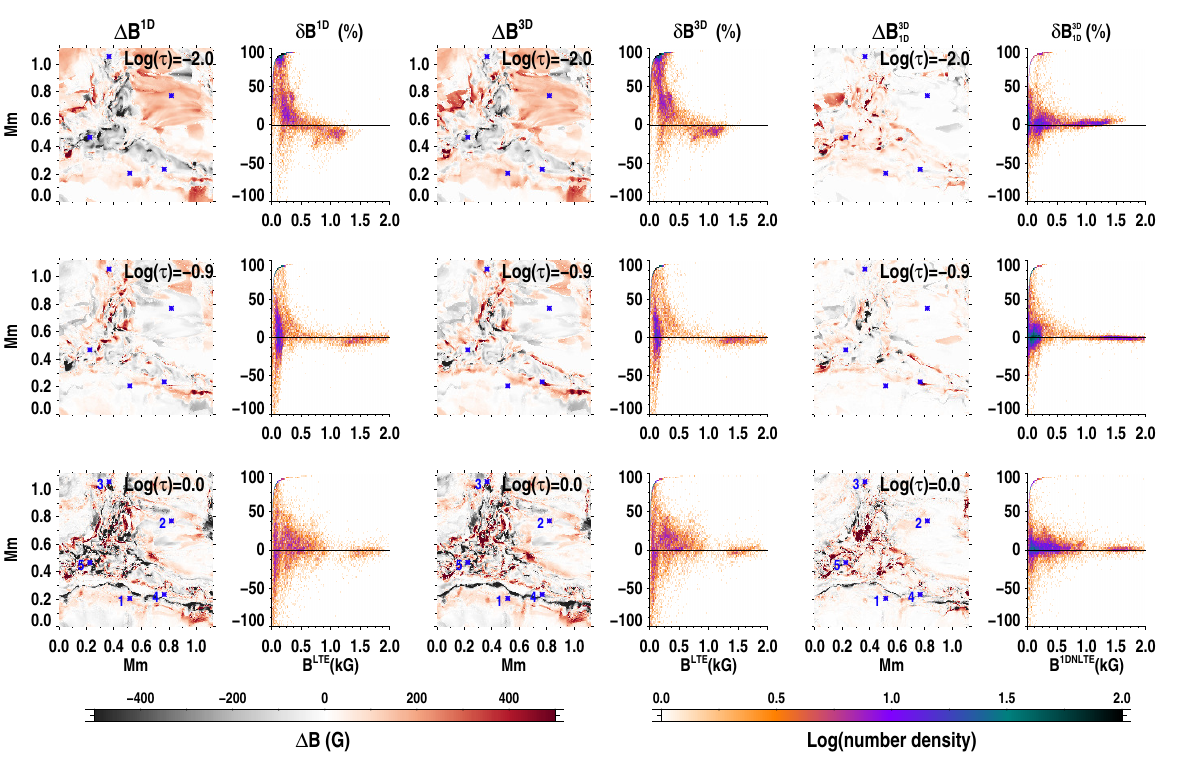}
\caption{Same as Figure~\ref{fig:temp} but for the magnetic field strength. See Section~\ref{sec:bfield} for more details.}
\label{fig:bfield}
\end{figure*}

\subsection{Line-of-sight velocity}
\label{sec:vlos}
We now discuss how the LOS velocity in the testmodel-3D deviates from the reference model. From the simple difference and relative difference maps shown in Figure~\ref{fig:vlos}, the errors in the testmodel-3D are of similar magnitude as in the testmodel-1D. These errors are concentrated at the boundaries of the granules, and along the intersection of the intergranular lanes and the magnetic structure. In these regions, the plasma is highly turbulent with strong velocity gradients. The 1DNLTE effects are more dominant than the 3D effects. We also see that in a fraction of the pixels, the relative differences are  large (>50\%). These come from regions of smaller velocities in the MHD cube. But the absolute error is more or less independent of the actual value. Like in paper I, we again neglect pixels with $v_{\rm LOS} < 10$\,m/s to avoid smaller denominators in Equation~\ref{eqn:rel_diff1}. The mean and standard deviations of the histograms $\Delta v^{\rm 1D}$ and $\Delta v^{\rm 3D}$ given in Table~\ref{tab:table1} are comparable, {while these values for $\Delta v^{\rm 3D}_{\rm 1D}$ are a few times 10\,m/s.}

The errors in the measurement of $v_{\rm LOS}$ at the five representative spatial positions in testmodel-1D and testmodel-3D (see Figure~\ref{fig:1d-3d}) are of similar magnitude. At Pos.\,1 and Pos.\,2, velocities in testmodel-1D and testmodel-3D match closely with the reference values. In the intergranular lane, at pos.\,3, errors are large in both the testmodels at $log(\tau)=-0.9$ and $0.0$. Here, a Doppler shift in the line core can be clearly seen in Figure~\ref{fig:profiles}. The velocity at Pos.\, 4. is quite interesting, because in testmodel-1D its value is close to the reference value, but in the testmodel-3D the error has increased significantly at all three nodes. As discussed in paper I, this is a point in the intergranular lane next to a magnetic structure. It is in a region of extreme gradients in flow velocities due to the tilting of magnetic flux sheet into the adjacent granule \citep[see Figure~10 of][]{2015A&A...582A.101H}. In this case we see that the 3D effects further strengthen the 1DNLTE effects resulting in a large deviation of the measured value in the testmodel-3D. In Pos.\,5, which is inside the magnetic structure sandwiched between cooler environments, we observe the opposite. That is, the 3D effect has weakened the 1DNLTE effect, thus reducing the error in the measured velocity, similar to the behaviour noticed for the temperature. 

 \citet{2013A&A...558A..20H} found that the Doppler shifts in the 1DNLTE profiles differ from the 3DNLTE profiles only by about 100 m/s. When we compare the velocity maps from the two test models (last two columns in Figure~\ref{fig:vlos}), the overall influence of 3D effects on the velocity determination is quite small. 3D RT effects are mostly observed at the boundaries of intergranular lanes and magnetic structures, which also happen to be the regions of strong 1DNLTE effects. The histograms of the $\Delta^{\rm 3D}_{\rm 1D}$ maps peak at zero in all three nodes with a standard deviation below $40$\,m/s. 

\begin{figure*}
\centering
\includegraphics[width=\textwidth]{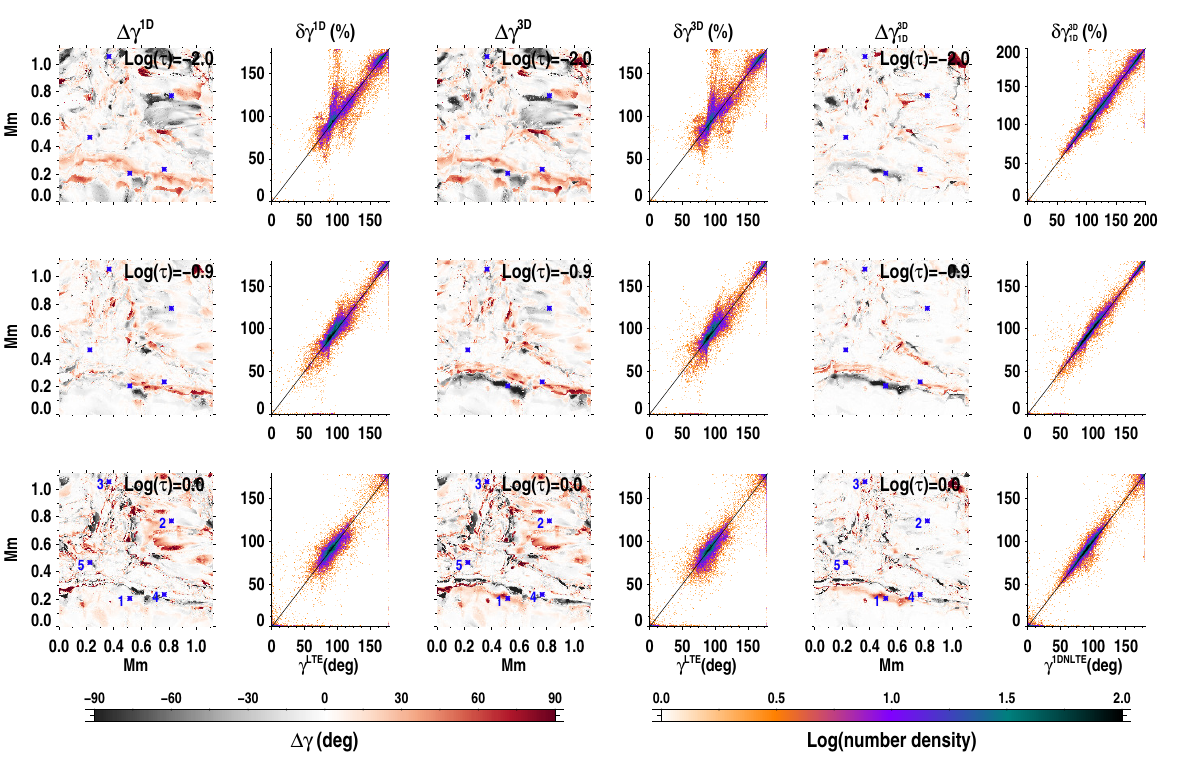}
\caption{Same as Figure~\ref{fig:temp} but for the magnetic field inclination. {See Section~\ref{sec:bfield} for more details.} }
\label{fig:gamma}
\end{figure*}    

\subsection{Magnetic field}
\label{sec:bfield}
Based on the calculations in a flux tube model, \cite{1978A&A....67...33S} noted that multi-dimensional RT effects can influence the magnetic field measurements. This was further corroborated by the findings of \cite{2012A&A...547A..46H}. We find this to be true also in an MHD cube. However statistically, the general properties of the difference maps $\Delta B^{\rm 1D}$ and $\Delta B^{\rm 3D}$ are similar, like in the case of temperature and velocity measurements. This is the case also for the magnetic field inclination and its difference maps plotted in Figure~\ref{fig:gamma}. From Figure~\ref{fig:bfield}, the relative difference $\delta B^{\rm 3D}$ is as high as 50\% or more for smaller values of $B^{\rm LTE} < 300\,$G, similar to $\delta B^{\rm 1D}$. {But in a majority of the pixels, the relative difference is within $20\%$}. The errors in testmodel-3D are smaller in the central node as it is best constrained in the inversions. A part of the errors at $log(\tau)=0$ are due to the difficulties in measuring small-scale weak fields close to the surface (notice the larger noise in panels from last rows of Figs. 6 and 7). However, the presence of clear structure at $log(\tau)=0.0$ in the diff between testmodels 1D and 3D, just as for other nodes, suggests that these difference {could be} due to 1DNLTE and horizontal transfer effects.

When we focus on the 3D effects in particular, differences between the testmodel-1D and testmodel-3D are bigger in the top and bottom nodes, and much less in the central node. In the top node, the 3D effects show up in pixels along the magnetic structure and a patch on the upper left which is at the intersection of a granule, an intergranular lane and a magnetic structure. The polarization profiles show variations due to 3D effects in this region (not shown in the paper) and this influences the inverted field strength. In addition, due to the different temperatures of the plasma within this small junction, we can expect the horizontal irradiation to affect the equivalent widths of the lines and thus affect the magnetic field strength. We also see the 3D effects around this region in the derived temperature maps of Figure~\ref{fig:temp}. The $\delta B^{\rm 3D}_{\rm 1D}$ maps in the top node suggest that the overall field strength in testmodel-3D is lower than in testmodel-1D. 

\begin{figure*}[htbp]
\centering
\includegraphics[width=\textwidth]{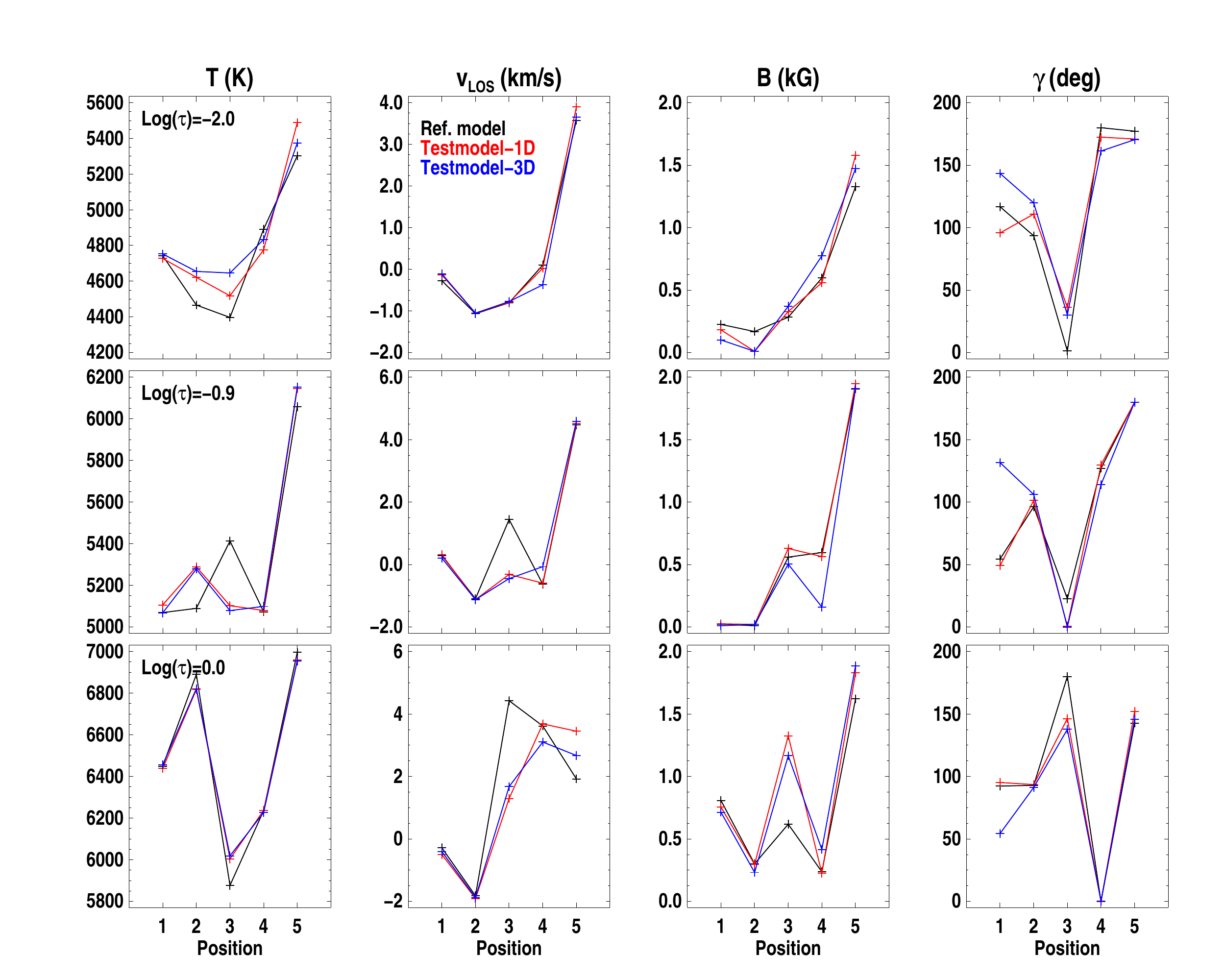}
\caption{Values of the different atmospheric quantities in the reference model (\textit{black}), testmodel-1D (\textit{red}) and in testmodel-3D (\textit{blue}) at the five representative spatial position, at all three inversion nodes. }
\label{fig:1d-3d}
\end{figure*}

When we consider the individual profiles, the 3D effects are seen to not only affect the intensity but also the polarization, albeit to a much smaller extent. {When the polarization signals are small, isolating the effects of 3D radiative transfer from the uncertainties in the inversions becomes difficult.} \cite{1978A&A....67...33S} and \cite{2012A&A...547A..46H} found that the 3D RT effects  weaken the $Q/I_c$ and $V/I_c$ profiles. Of all the 5 positions in our selection, the $V$ profiles at pos.\,1 show {noticeable} weakening by 3D effects. {Although small differences in the 3DNLTE case are seen also in linear polarization, the profiles are quite weak (<1\%).} When we examine the inclination maps in the reference model around pos.\, 1 (Figure~\ref{fig:maptau}), the inclination changes quite drastically. This can contribute to the horizontal transfer effects in the polarization profiles. {We see differences as large as $50^\degree$ between the inclinations found in testmodel-1D and testmodel-3D in pos.\,1 and the region surrounding it, at all the three nodes (Figures~\ref{fig:gamma}, \ref{fig:1d-3d}). Since such a large difference is not specific to one isolated pixel but is observed over an extended region following the boundary between the granule-intergranular lane, it is possible that the horizontal transfer effects are playing a role.} In particular, the 3D effects at this position have shifted the measured inclination further away from the reference value. 

At pos.\,5, we see the exact opposite as all the Stokes profiles, $I/I_c, Q/I_c, U/I_c$ and $V/I_c$ show strengthening from the 3D effects. This supports the findings of \cite{2015A&A...582A.101H} that the horizontal RT strengthens the line at the centers of magnetic elements. The effect is especially clear in the 6301.5\,\AA{} line. In the second line at 6302.5\,\AA{}, the 3D effects weaken the $\pi$ components and strengthen the blue $\sigma$ components in both $Q/I_c$ and $U/I_c$ profiles. This is because of the regions with different temperatures surrounding pos.\,5.  The magnetic field strength at $log(\tau)=-2.0$ in testmodel-3D has shifted closer to the reference value but away from it in the bottom node. 

At the other three positions, pos.\,2, pos.\,3 and pos.\,4, the polarization profiles computed in 3DNLTE are slightly weaker than the 1DNLTE profiles (Figure~\ref{fig:profiles2}). The errors due to the neglect of 3D effects are at some nodes enhanced and in others, reduced (Figure~\ref{fig:1d-3d}).

\begin{figure*}
\centering
\includegraphics[width=\textwidth]{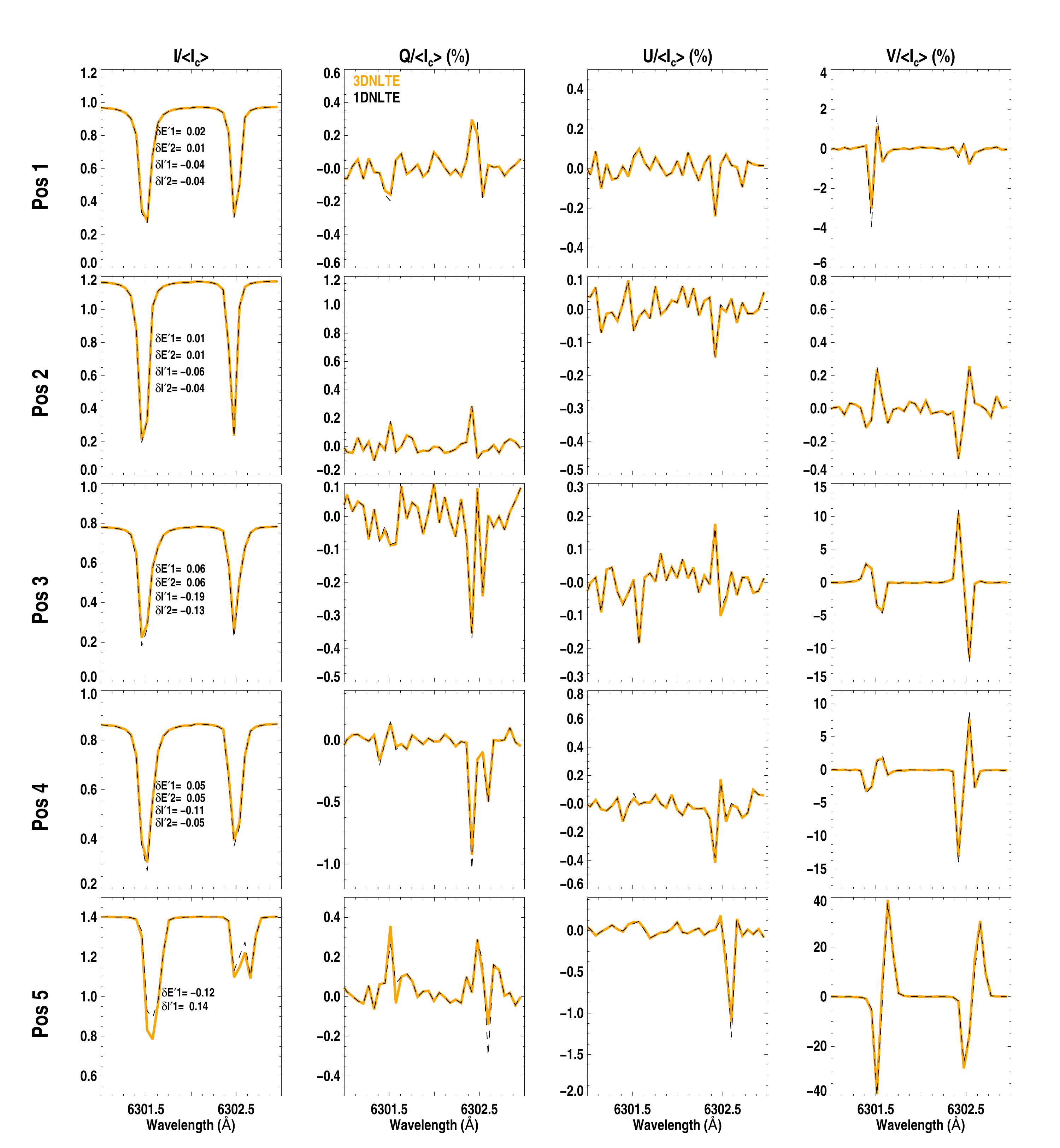}
\caption{Figure showing the effects of spectral degradation and noise on the Stokes profiles computed in 1DNLTE (black) and 3DNLTE (orange), at the five spatial positions. The differences in equivalent widths $\delta E^{\prime}$ and residual intensities $\delta I^{\prime}$ are computed using Equation~\ref{eqn:int_ew_2}. See Section~\ref{sec:deg_prof} for more details. }
\label{fig:prof_specmear}
\end{figure*} 

\begin{figure*}
\centering
\includegraphics[width=0.9\textwidth]{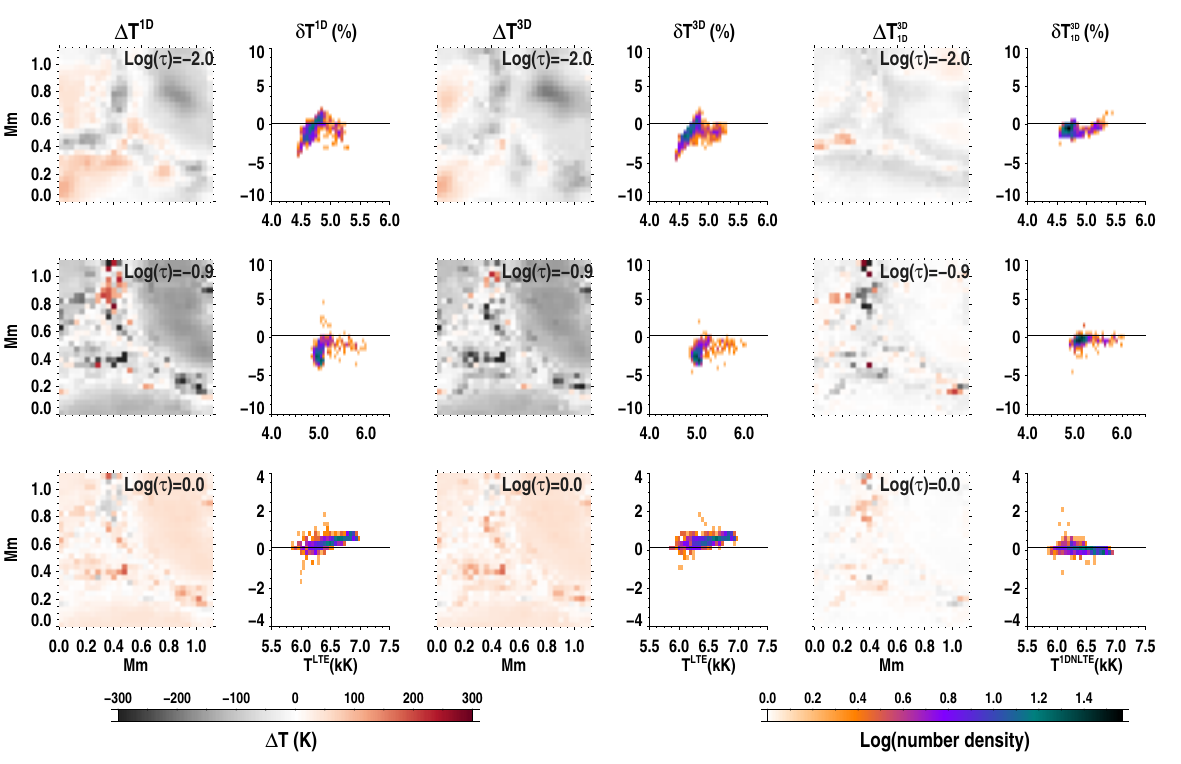}
\caption{Same as Figure~\ref{fig:temp} but for the Stokes profiles degraded to the specifications of the CRISP instrument on the Swedish Solar Telescope (SST). See Section~\ref{sec:deg_prof} for more details.}
\label{fig:temp_deg}
\end{figure*} 

\section{Effects of spatial and spectral degradation}
\label{sec:deg_prof}

In paper I, we showed that the 1DNLTE effects in the Stokes profiles survive even when they are spatially averaged. In the present paper, we have conducted a similar test to check the strength of  3DNLTE effects when the Stokes profiles are degraded both spatially and spectrally. For this, we consider the specifications of the CRisp Imaging SpectroPolarimeter \citep[CRISP,][]{2008ApJ...689L..69S} at the Swedish Solar Telescope \citep[SST,][]{2003SPIE.4853..341S}. The data were degraded to a spectral resolution of 60\,m\AA{}, and a spatial resolution of $0.1^{\arcsec}$. The spectral and spatial point spread functions (PSF) were assumed to be Gaussian. In addition, we added a Gaussian noise with a standard deviation of 1x10$^{-3}I_{c}$. The inversions were done in a similar way to the case without any degradation (Section~\ref{sce:inversion}). 

Figure~\ref{fig:prof_specmear} shows the effects of spectral smearing and noise on the Stokes profiles from Figure~\ref{fig:profiles2}. Due to their weak signals, the polarization profiles are more affected by such a spectral degradation than the intensity profiles. The differences between the 1DNLTE and 3DNLTE profiles, although small in the non-degraded case, are still observed after spectral smearing and wavelength rebinning. The values of $\delta E^{\prime}$ and $\delta I^{\prime}$, computed using Equation~\ref{eqn:int_ew_2}, are close to the values measured for the original profiles. In a few cases such as at Pos.\,1 and Pos.\,2, spectral degradation has  reduced $\delta I^{\prime}1$ by a small amount.

The temperature in the atmospheric models obtained by the inversion of spatially and spectrally degraded profiles are compared in Figure~\ref{fig:temp_deg}. We see a similarity between Figures~\ref{fig:temp_deg} and \ref{fig:temp}, both in the simple difference and in relative difference maps. While in some regions, the spatial and spectral degradation has reduced the 3DNLTE effects, in some other regions, we see differences much larger than before. The latter could be a result of spatial binning. This is because, such an effect was also seen in paper I (see Figure~12 of that paper) where the profiles were only spatially binned and no other degradation was applied. The spatial smearing and noise can also further enhance this effect.  The LOS velocity and magnetic field maps shown in Appendix~\ref{sec:appendix-b} also show similar behaviour. 

We repeated this test by degrading the data to the specifications of the  Visible Spectro-Polarimeter \citep[ViSP,][]{visp} instrument at the Daniel K. Inouye Solar Telescope (DKIST). Since the ViSP  observes at a much higher resolution that the CRISP, the effects of 1DNLTE and horizontal transfer were seen at close to the same level as in the original simulation.

\section{Summary and conclusions}
\label{sec:conclusions}
In continuation to paper I, we here study the errors introduced in the inverted model atmosphere when 3D radiative transfer effects are neglected. We selected a small region of the large MHD cube used in paper I to perform the 3DNLTE computations and then invert the profiles using the LTE inversion code, SPINOR. The derived atmosphere, called testmodel-3D, is then compared with the reference model obtained from the inversion of LTE profiles. Any departure between the two is attributed to 3DNLTE effects. In addition to this, we also compare the atmospheric model from the inversion of 1DNLTE profiles called the testmodel-1D, over the same region. The testmodel-1D and testmodel-3D are compared to isolate the 3D RT effects. 

 \cite{2013A&A...558A..20H, 2015A&A...582A.101H} nicely summarized that the 3D effects weaken the lines when there is horizontal irradiation from hotter surroundings, and it strengthens the line profile when the surroundings are cooler. It is the derived quantities in such regions that show departures in testmodel-3D compared to testmodel-1D. This is observed along the boundaries of magnetic structures and their surrounding intergranular lanes, granules-intergranular lane boundaries, and in pixels inside magnetic structures affected by irradiation from cool intergranular lanes.
Since the 3D effects can weaken or strengthen the 1DNLTE line profiles, it can either enhance or reduce the departure from the reference model. 

\begin{figure*}
\centering
\includegraphics[width=0.9\textwidth]{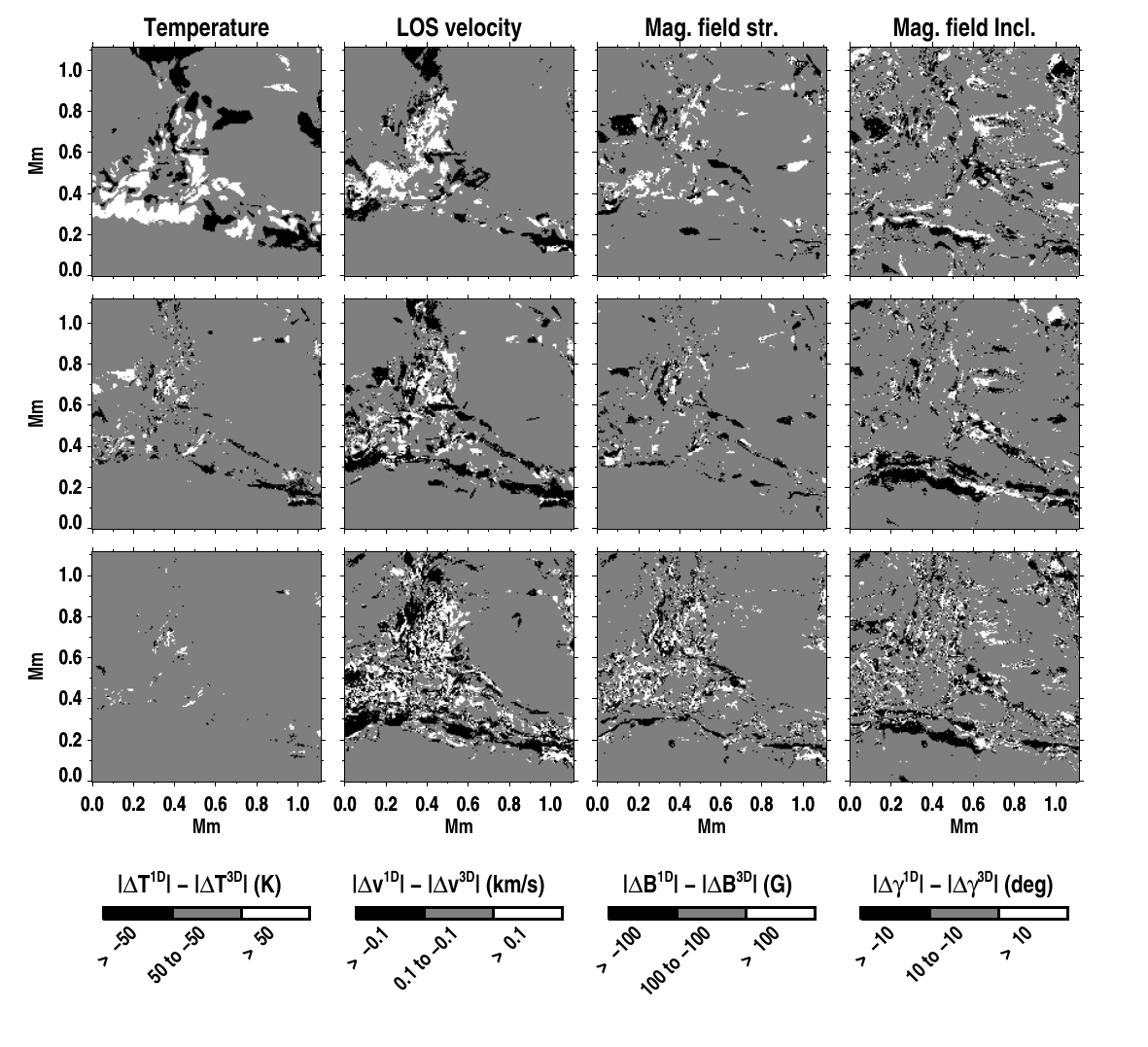}
\caption{Patches representing the regions where horizontal RT enhances the NLTE effects (\textit{white}), weaken the NLTE effects (\textit{black}), or has no significant influence (\textit{grey}). This is shown for, from left to right, $T, v_{\rm LOS}, B$ and $\gamma$ at all the three inversion nodes; from top to bottom $log(\tau)=-2.0, -0.9, 0.0$. The threshold values employed to divide the regions into the three categories are indicated in the colorbar. {See Section~\ref{sec:conclusions}} for further details.}
\label{fig:bmap}
\end{figure*}

  Figure~\ref{fig:bmap} demonstrates where in the sub-region of the cube, the 3D effects enhance or weaken the 1DNLTE effects. Pixels in black highlight the regions where 3D effects enhance 1DNLTE effects, that is, the values in testmodel-3D are further shifted away from the reference model compared to those in testmodel-1D. Regions highlighted in white show the opposite effect, they are the regions where the 3D effects weaken the 1DNLTE effects. Regions in grey represent pixels where the 3D effects are weak and so do not fall into either of the two categories. The threshold values for dividing the pixels into the above described three categories are $50$\,K, $100$\, m/s, $100$\,G and $10^\degree$ in $T, v_{\rm LOs}, B$ and $\gamma$, respectively. 

It can be seen from this figure that, the 3D effects are most prominent in and around the magnetic structure and they clearly affect the temperature and velocity measurements. Since the 3D effects mostly affect the line core, we see their influence more clearly in the top node. 
Of the regions that are affected (black and white patches), the departures range between 50\,K -- 100\,K. 

In velocity maps, the influence of 3D effects can alter the values between 100\,m/s -- 500\,m/s. We also see some effects in the central and bottom nodes, in regions of strong velocity gradients, such as at pos.\,4 shown in Section~\ref{sec:vlos} and discussed in paper I. These gradients are difficult to recover in the simple 3-node atmospheric model used by us. 

In the case of magnetic field strength and inclination, the white and black patches are scattered all across the sub-region. The errors in the magnetic field strength due to the neglect of 3D effects is in the range 100\,G -- 500\,G and in inclination, it is between $10^\degree$ -- $50^\degree$.  We see significant influence of 3D effects on the measured magnetic field inclination along a patch below a magnetic structure, as discussed in Section~\ref{sec:bfield}. Only in this patch, the errors are larger than $50^\degree$ in a few pixels. 

To conclude, the errors introduced in the inverted atmosphere due to the neglect of 3D RT effects are not as large as the errors when NLTE effects are neglected completely, at least over the sub-region analysed here. The 3D effects are more localised to regions surrounded by strong horizontal gradients in temperature and intensity, such as magnetic flux concentrations \citep{Carlsson_2004}. Neglecting their effects not only introduces errors in temperature determination but, to a lesser extent, also in other parameters. In Section~\ref{sec:deg_prof}, we have tested how the spatial and spectral degradation of the Stokes profiles in the presence of noise, affects the 1DNLTE and horizontal radiative transfer effects. {The difference maps comparing the reference and testmodels show a similarity to the case without degradation.} This demonstrates that even at low spectral and spatial resolutions, the statistical nature of the errors introduced in the inverted atmosphere by the neglect of 1DNLTE and/or 3D effects can remain the same. 

The 3D effects can be expected at granular boundaries, intergranular lanes and magnetic bright points. In the construction of atmospheric models for photospheric bright points \citep[e.g.,][]{Shelyag2010, Hewitt2014}, the 3D effects are neglected although they can affect the temperature stratification. Given the results of this study, we expect that the 3DRT effects are also important in other photospheric features such as sunspots where we observe horizontal gradients in temperatures across the umbra-penumbra boundaries, penumbral filaments and spines, boundaries of light bridges, and umbral dots. The atmospheric diagnostics of these features \citep{Socas_Navarro_2004, Riethm_ller_2008, 2013A&A...557A..25T, 2015A&A...583A.119T} carried out by neglecting 3D effects can result in enhanced errors in the deduced quantities. 

Currently there are no inversion codes capable of performing 3DNLTE calculations because it requires large computational resources. However, it is important to consider the errors that can get introduced in the inverted atmosphere when 3D radiative transfer effects are neglected.

\begin{acknowledgements}
{We thank the anonymous referee for constructive comments which helped improve the paper.} This project has received funding from the European Research Council (ERC) under the European Union's Horizon 2020 research and innovation programme (grant agreement No. 695075). This work has also been partially supported by the BK21 plus program through the National Research Foundation (NRF) funded by the Ministry of Education of Korea. This research has made use of NASA’s Astrophysics Data System.
\end{acknowledgements}


\begin{appendix}
\section{Testmodels and fit to the Stokes profiles}
\label{sec:appendix-a}
\begin{figure*}[h!]
    \centering
    \includegraphics[width=0.8\textwidth]{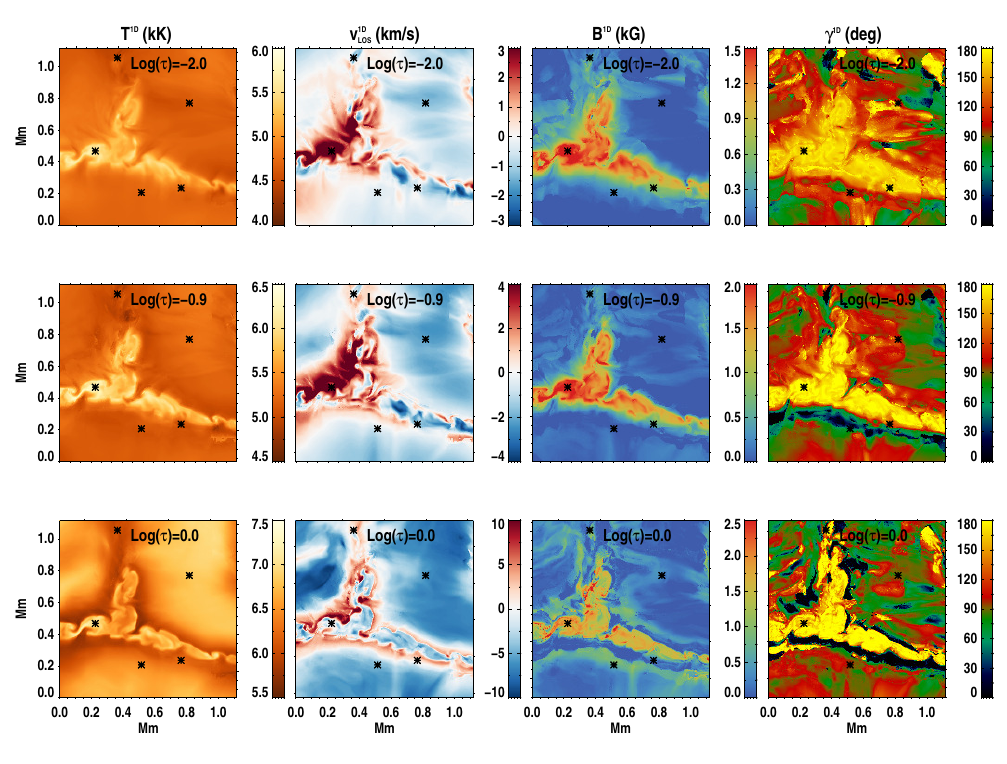}
    \caption{Maps of different atmospheric quantities in testmodel-1D.}
    \label{fig:model-1d}
\end{figure*}
\begin{figure*}[h!]
    \centering
    \includegraphics[width=0.8\textwidth]{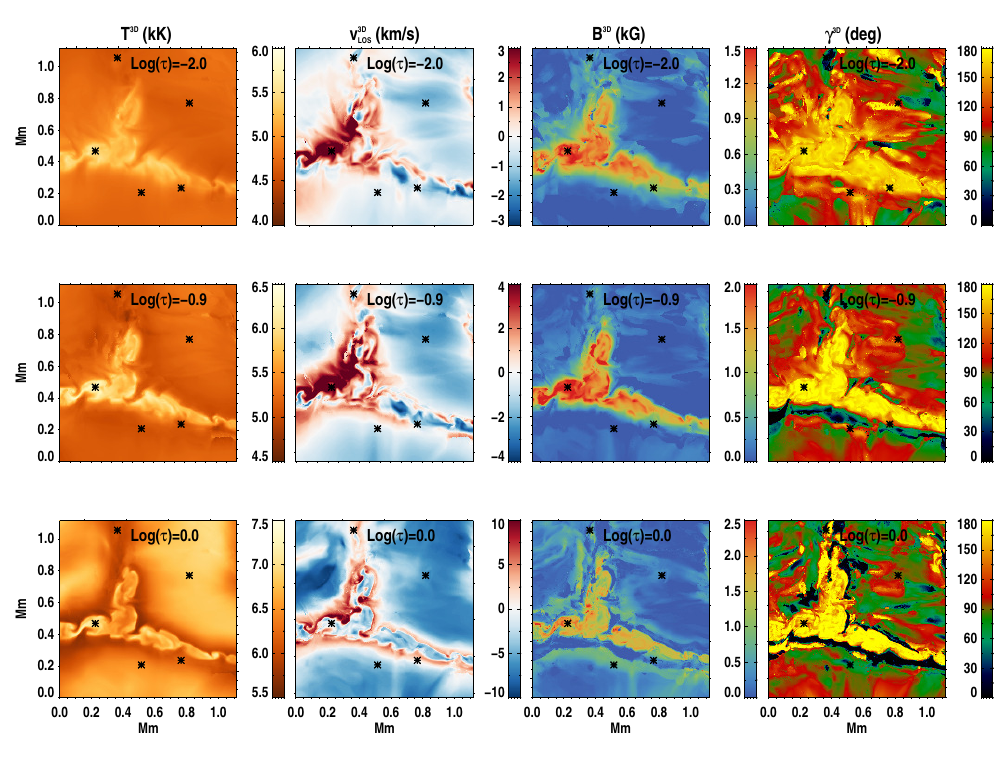}
    \caption{Maps of different atmospheric quantities in testmodel-3D.}
    \label{fig:model-3d}
\end{figure*}
\begin{figure*}[h!]
    \centering
    \includegraphics[width=0.8\textwidth]{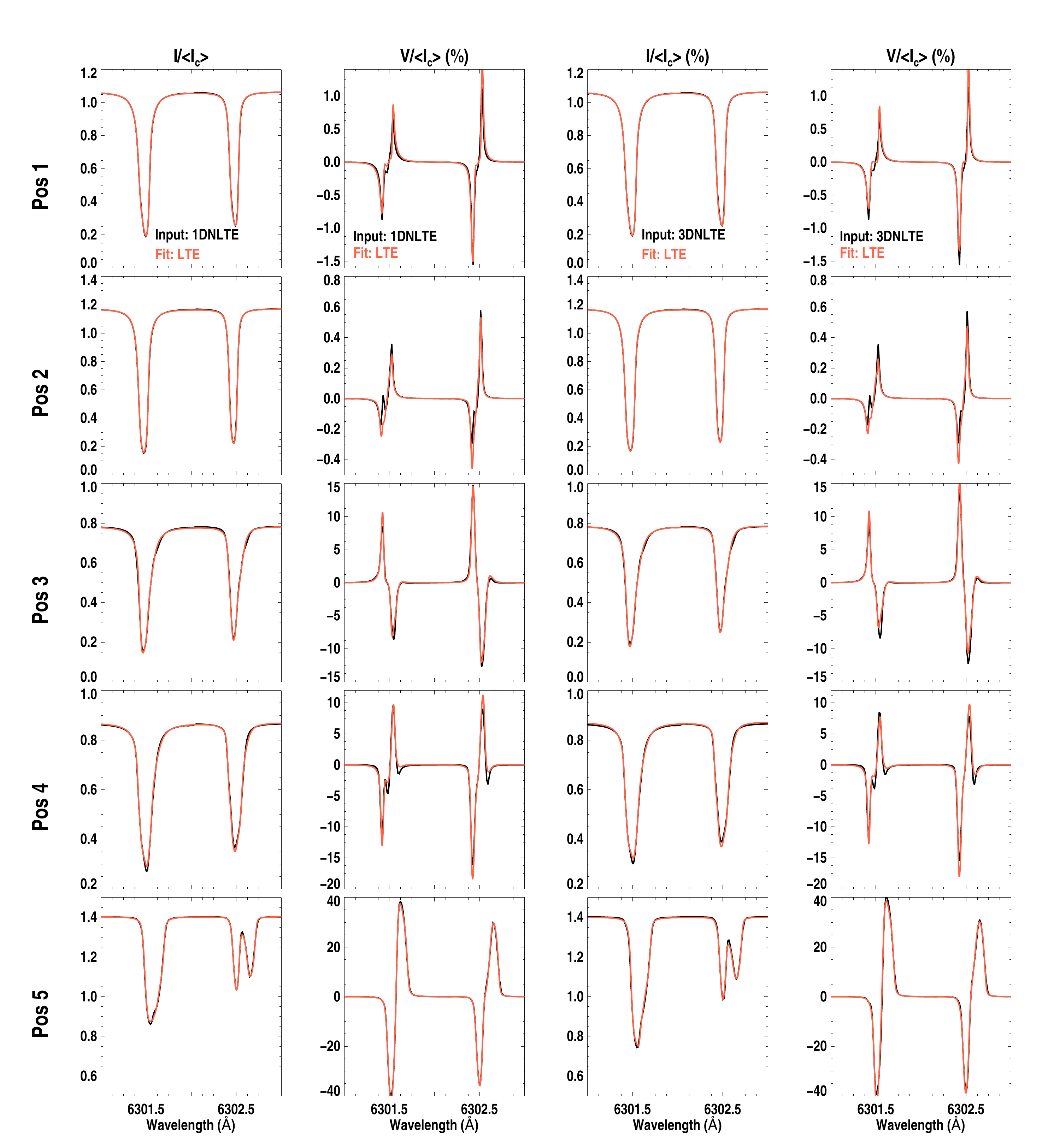}
    \caption{{LTE fit to the Stokes profiles shown in Figure~\ref{fig:profiles2}. The input profiles are computed in 1DNLTE (\textit{first two columns}) and in 3DNLTE (\textit{last two columns}). }}
    \label{fig:fit_profiles}
\end{figure*}

\section{Effects of spatial and spectral degradation on the velocity and magnetic field measurements}
\label{sec:appendix-b}
\begin{figure*}[h!]
    \centering
    \includegraphics[width=0.8\textwidth]{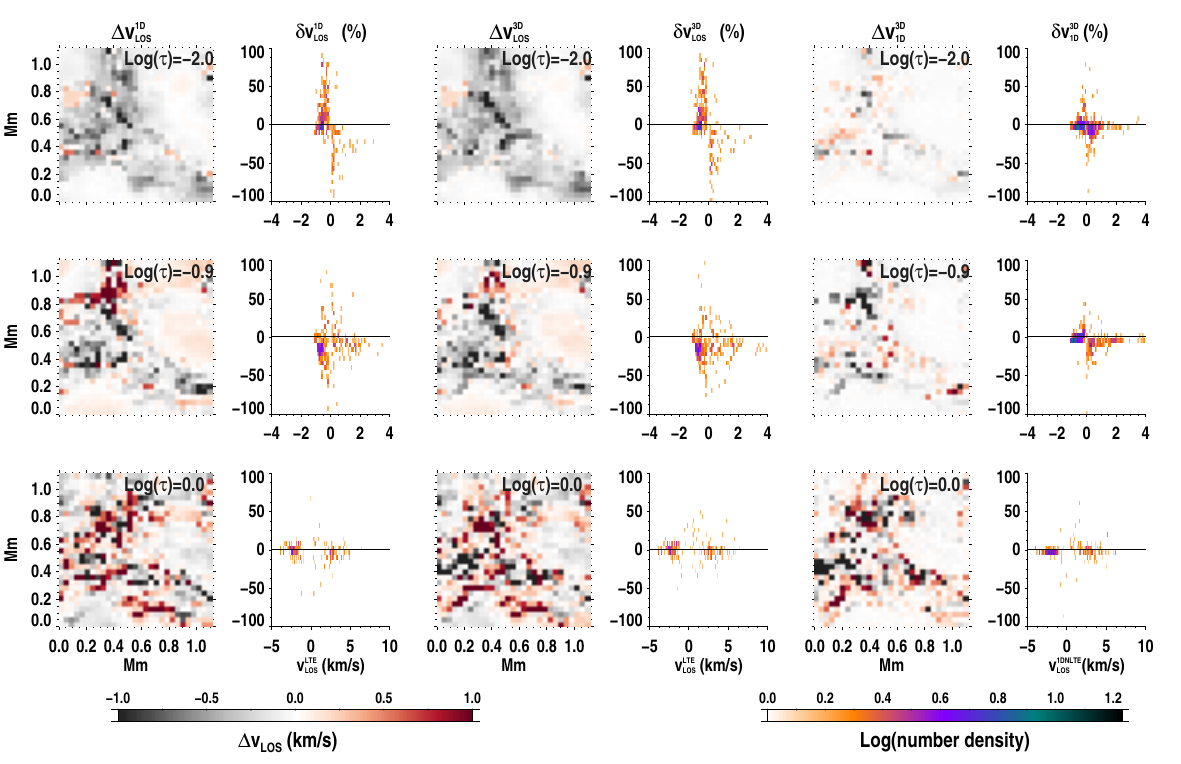}
    \caption{Same as Figure~\ref{fig:vlos} but for the Stokes profiles degraded to the specifications of the CRISP instrument on the Swedish Solar Telescope (SST). See Section~\ref{sec:deg_prof} for more details.}
    \label{fig:vlos-deg}
\end{figure*}

\begin{figure*}[h!]
    \centering
    \includegraphics[width=0.8\textwidth]{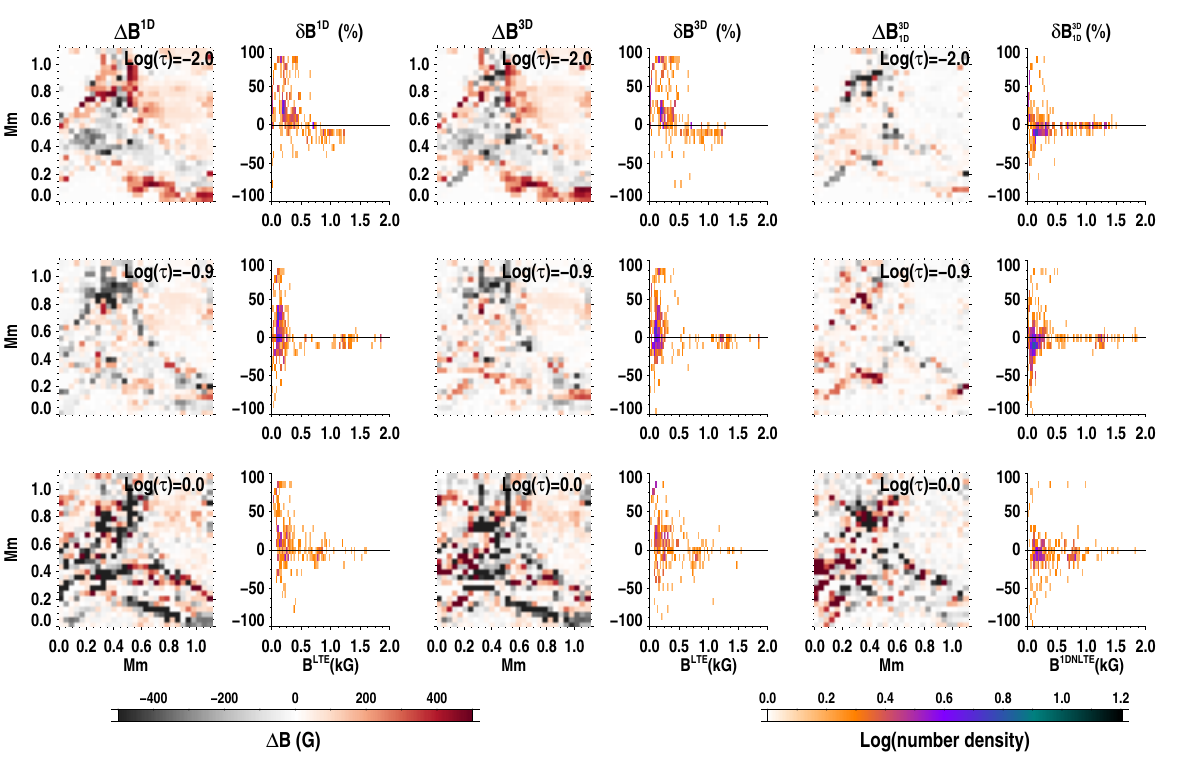}
    \caption{Same as Figure~\ref{fig:bfield} but for the Stokes profiles degraded to the specifications of the CRISP instrument on the Swedish Solar Telescope (SST). See Section~\ref{sec:deg_prof} for more details.}
    \label{fig:bfield-deg}
\end{figure*}

\begin{figure*}[h!]
    \centering
    \includegraphics[width=0.8\textwidth]{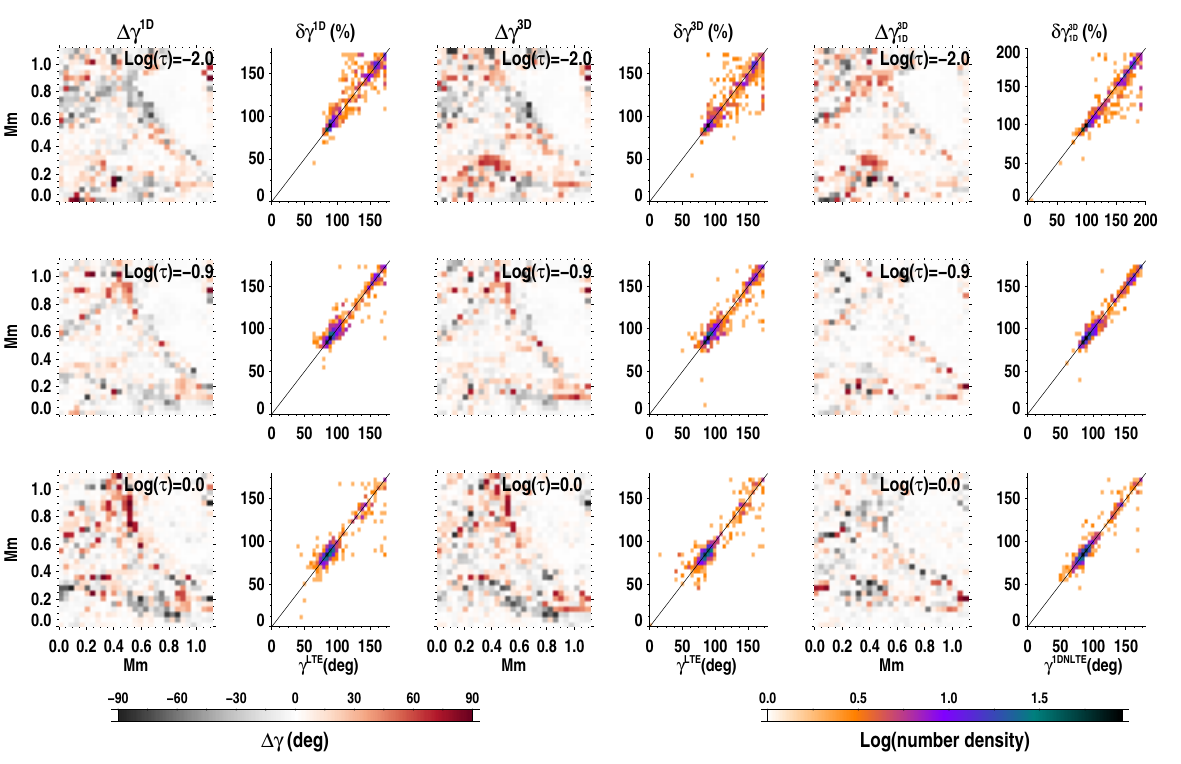}
    \caption{Same as Figure~\ref{fig:gamma} but for the Stokes profiles degraded to the specifications of the CRISP instrument on the Swedish Solar Telescope (SST). See Section~\ref{sec:deg_prof} for more details.}
    \label{fig:gamma-deg}
\end{figure*}

\end{appendix}

\end{document}